\documentclass[12pt]{article}
\usepackage{graphicx}
\usepackage{amsmath}
\usepackage{amsfonts}
\usepackage{amssymb}
\newtheorem{theorem}{Theorem}[section]

\newtheorem{lemma}[theorem]{Lemma}

\numberwithin{equation}{section}

\begin{document}

\title{Sharp asymptotics for Einstein-$\lambda$-dust flows}

\author{
Helmut Friedrich\\ 
Max-Planck-Institut f\"ur Gravitationsphysik\\
Am M\"uhlenberg 1\\
14476 Golm, Germany}

\maketitle

{\footnotesize

\begin{abstract}
We consider the Einstein-dust equations  with positive cosmological constant $\lambda$ on manifolds with time slices diffeomorphic to an orientable, compact 3-manifold
$S$. It is shown
that the set of standard Cauchy data 
for the Einstein-$\lambda$-dust equations on $S$
contains an open (in terms of suitable Sobolev norms) subset 
of data that develop into solutions which admit  at future time-like infinity 
a space-like conformal boundary ${\cal J}^+$ that is $C^{\infty}$ if the data are of class $C^{\infty}$ and of correspondingly lower smoothness otherwise.
As a particular case follows a strong  stability result for FLRW solutions.
 The solutions can conveniently be characterized in terms of their asymptotic end data  induced on ${\cal J}^+$, only a linear equation must be solved to construct such data. In the case where the energy density $\hat{\rho}$ is everywhere positive
 such data can be constructed without solving any differential equation at all.

\end{abstract}


\newpage

\section{Introduction}

It has been known for a while that among the solutions to Einstein's vacuum field equations $\hat{R}_{\mu \nu} = \lambda\,\hat{g}_{\mu \nu}$
with positive cosmological constant $\lambda$ on manifolds with space-sections diffeomorphic
to an orientable, compact  3-manifold
 $S$ there is an open (in terms of Sobolev norms on Cauchy data) subset of solutions which are future asymptotically simple in the sense of Penrose  \cite{penrose:1965}, i.e. the solutions admit the construction of a conformal boundary ${\cal J}^+$ at their infinite time-like future which is $C^{\infty}$ if the solutions are $C^{\infty}$  and is of correspondingly lower smoothness otherwise (see  \cite{friedrich:beyond:2015} for more details and references). This property generalises to the Einstein-$\lambda$ equations coupled to conformally covariant matter field equations with trace free energy momentum tensor.  In \cite{friedrich:1991} this has been discussed in detail
 for the  Maxwell and the Yang-Mills equations, where a procedure has been laid out which applies, possibly which some modifications in specific cases, to other such field equations (see \cite{luebbe:valiente-kroon:2013} for a recent example).

Matter fields with energy momentum tensors which are  not trace free were generally expected to lead to difficulties in the construction of reasonably smooth conformal boundaries.
(The emphasis here is on results about the evolution problem, we are not talking about geometric studies near conformal boundaries which postulate  properties  of energy momentum tensors convenient for their analysis).
It has recently been observed, however, that this need not be the true \cite{friedrich:massive fields}.

In the case of the Einstein-Klein-Gordon equations the conformal field equations with suitably transformed matter field  imply evolutions system which are hyperbolic, irrespective of the sign of the conformal factor, if the mass and the cosmological constants are related by the equation $m^2 = \frac{2}{3}\,\lambda$.  If this condition is imposed a fairly direct calculation shows that the equation for the rescaled scalar field becomes regular where the conformal factor goes to zero. However, that the conformal equations for the geometric fields become regular in this limit is far from immediate and, as in the case discussed in the following, came as a surprise after various attempts to cast the singular equations into a form which would allow one to draw conclusions about the precise asymptotic behaviour of the solutions
in the presence of singularities.

Leaving aside the questions about the significance of this particular result, the present article is concerned with the analysis of another matter model with non-vanishing trace of the energy momentum tensor. We study in detail the future asymptotic behaviour of solutions to the Einstein-$\lambda$-dust equations.

In a recent article Had$\check{z}$i\'c and  Speck have shown that the FLRW solutions
to the Einstein-$\lambda$-dust equations  with underlying manifolds of the form $\mathbb{R} \times \mathbb{T}^3$ are future stable, i.e. slightly perturbed FLRW data on $ \mathbb{T}^3$ develop into solutions to the Einstein-$\lambda$-dust equations whose causal geodesics are future complete \cite{hadzic-speck-2015}. The authors use the method proposed in \cite{friedrich:hyperbolic} to control the evolution of a general wave gauge in terms of its  gauge source functions. As emphasized in \cite{friedrich:hyperbolic}, it is clear that (under fairly weak smoothness assumptions) any coordinate system can in principle be controlled in terms of its gauge source functions and suitable initial data. But finding gauge source  functions which are useful  in a specific problem is quite a delicate matter. 
The authors manage to  identify gauge source functions which allow them to derive estimates that give  control on the long time evolution of their solutions (see \cite{ringstroem:2008} for another such case).

It is, however, quite a different question whether the gauge so established lends itself to analyzing the asymptotic behaviour of solutions in detail and to deciding, for instance, whether the differentiable as well as  the conformal structure of the solutions  admit simultaneously extensions of some smoothness to (future) time-like infinity as required by asymptotic simplicity.

FLRW solutions are known to be future asymptotically simple (see section \ref{FLRW-sols}).
This may be expected to be is just an artifact of the high symmetry 
requirements which imply local conformal flatness and hypersurface orthogonality of the flow field.
The present study grew out of attempts to understand what may go wrong under  more general assumptions and what kind of obstruction to the asymptotic smoothness of the conformal structure may possibly arise from the presence of a non-vanishing energy density $\hat{\rho}$.

In the article \cite{friedrich98} have been derived hyperbolic evolution equations from the Einstein-dust equation in a geometric gauge based of the flow field. The following analysis may be seen as a conformal version of this discussion. After presenting the Einstein-$\lambda$-dust equations in section \ref{Einstein-lambda-dust}, we derive in section \ref{conf-field-equ} the conformal field equations and suitably transformed matter field equations. It turns out that two equations of the system are singular in the sense that there occur factors of the form $\Omega^{-1}$ on the right hand side, where $\Omega$ is the conformal factor which is positive on the {\it physical} solution space-time and relates the {\it physical} metric $\hat{g}_{\mu\nu}$ there to the {\it conformal} metric $g_{\mu\nu}$ by $g_{\mu\nu} = \Omega^2\,\hat{g}_{\mu\nu}$. Since things are to be arranged such that 
$\Omega \rightarrow 0$ at future time-like infinity, where we want to understand the precise nature of the solutions, there arise problems. One of the singularities, namely the one in the transformed (geodesic) flow field equation, was to be expected. Much more serious is a singularity in the equation for the rescaled conformal Weyl tensor 
$W^{\mu}\,_{\nu \lambda \rho} = \Omega^{-1}\,C^{\mu}\,_{\nu \lambda \rho}[g]$, which plays a central role in the system. The singularities carry, however, interesting geometric information.
They imply that the (so far formally defined) set $\{\Omega = 0\}$ can only define a smooth conformal boundary of the solution space-time if the flow lines approach this set orthogonally.
Thus, if one wants to approach the problem in terms of estimates, one has to aim for sufficient  control to be able to define simultaneously a conformal boundary at time-like infinity, if admitted by the solution at all, and correspondingly control the  behaviour of the flow lines.

In the present article we try to exploit the  conformal properties of the system in the most direct way. In section \ref{the-reg-rel} it is shown that due to the specific form of the energy momentum tensor for dust the geodesics tangent to the flow field can be identified after a parameter transformation with curves underlying certain conformal geodesics. Since conformal geodesics are invariants of the conformal structure, this opens the possibility to define a gauge which extends regular across the conformal boundary ${\cal J}^+ = \{\Omega = 0\}$ if the latter can indeed be attached in a smooth way to the solution manifold (on which $\Omega > 0$, of course). It turns out that this gauge implies a certain {\it regularising relation} which proves useful in three different contexts. Its first important merit is to render the conformal field equations regular.
 
 In section \ref{hyp-red-equ} it is shown that the conformal field equations imply a 
{\it  hyperbolic reduced system of evolution equations} which can make sense up to and beyond the conformal boundary at time-like infinity (if it exists). This system is not obtained immediately. 
  The regularizing relation leads to a system which is hyperbolic where 
 $\Omega > 0$ but becomes singular where $\Omega \rightarrow 0$.
 A further regularization is performed to obtain a system
 which is hyperbolic independent of the sign of the conformal factor.

In section \ref{subs-syst} is derived a subsidiary system which implies that solutions to the hyperbolic evolution system for data that satisfy the constraints on a given Cauchy hypersurface (with respect to the metric provided by the evolution system) will satisfy in fact the complete system of conformal field equations. This closes the hyperbolic reduction argument.

To obtain complete information on the class of future asymptotically simple solutions
to the Einstein-$\lambda$-dust solutions we characterize in Lemma \ref{free-data-on-scri}
the possible {\it asymptotic end data} which may be prescribed on the conformal boundary 
${\cal J}^+ = \{\Omega = 0\}$ (assumed to be 3-dimensional, orientable, compact) of a solution that admits the construction of such a boundary with sufficient smoothness. As observed already
in \cite{friedrich:1986a} in the vacuum case,  the constraints reduce on ${\cal J}^+$  
 to a linear system of equations. Remarkably,  there is a case where the problem of solving the constraints simplifies even further. In the case where the density
 $\hat{\rho}$ is positive everywhere certain fields can be prescribed completely freely on  ${\cal J}^+$ and 
the rest follows by algebra and taking  derivatives. 
There is no need  to solve any differential equation at all 
(but see the remarks following Lemma \ref{free-data-on-scri}).

The reduced system of evolution equations is used in  section \ref{ex-strong-stab} to derive our main results. Being based on hyperbolic equations, a completely detailed statement of the results should give information about Sobolev norms. Since we only use properties of symmetric hyperbolic systems which can be found 
in the literature  at various places and because we are mainly interested in solutions of class $C^{\infty}$, we refrain from listing Sobolev indices. We would consider these only be of interest
if the weakest  possible smoothness assumptions were needed in the context of some concrete problems.

\begin{theorem}
\label{main-result}
Let $S$ be a smooth, orientable, compact 3-manifold, assume $\lambda > 0$, and denote by
${\cal A}_{\lambda, S}$ the set of standard Cauchy data on $S$  
to the Einstein-$\lambda$-dust equations with energy density $\hat{\rho} \ge 0$. Then

\vspace{.1cm}

\noindent
(i) There is an open (with respect to suitable Sobolev norms) subset ${\cal B}_{\lambda, S}$ of data in ${\cal A}_{\lambda, S}$ 
which develop into solutions that admit the construction of conformal boundaries in their infinite time-like future which are of class $C^{\infty}$ if the data are of class $C^{\infty}$ and of correspondingly lower differentiability if the data are of lower differentiability. 

\vspace{.1cm}

\noindent
(ii) The solutions which develop from data in  ${\cal B}_{\lambda, S}$ are completely parametrized by the asymptotic end data on $S$ (specified in Lemma \ref{free-data-on-scri}) which correspond to the data induced on the future conformal boundaries ${\cal J}^+$ of the solutions.

\end{theorem}

The case of the Nariai solution, an explicit, geodesically complete  solution to the Einstein-$\lambda$-dust equations 
with $\hat{\rho} = 0$ that admits not even a patch of a smooth conformal boundary  (see \cite{friedrich:beyond:2015}), shows that our reduced evolution system is by itself not sufficient to ensure the existence of a smooth conformal  boundary. Some extra information on the Cauchy data is required. 

Because the FLRW solutions do admit a smooth conformal future boundary one could consider data close to FLRW data. 
Following instead the arguments introduced in \cite{friedrich:1986b} and \cite{friedrich:1991},
a much larger class of suitable reference solutions (which includes the FLRW solutions)  
will be constructed in  section \ref{ex-strong-stab}  by solving a backward  Cauchy problem
for the reduced equations with asymptotic end data that are given on a 3-manifold $S$ which in the end will represent  the future conformal boundary ${\cal J}^+ = \{ \Omega = 0\}$ of the physical space-time defined on the set $\{\Omega > 0\}$.

In a second step we  consider  the `physical' standard Cauchy data that are induced by one of these solutions on a `physical' Cauchy hypersurface. It is shown that under sufficiently small perturbations of these data the resulting  solutions are {\it strongly stable} in the sense that  the smooth extensibility  of their conformal  structures at future time-like infinity is preserved. This makes use of the fact that a future asymptotically simple solution admits a conformal representation that extends as a smooth solution to the conformal Einstein-$\lambda$-dust  equations beyond the conformal boundary into a domain where $\Omega < 0$. The strong stability result follows then as a consequence of the well known Cauchy stability property of hyperbolic equations and the fact that the equation themselves ensure that the set of points where $\Omega = 0$ defines a smooth space-like hypersurface. 

Though they lead to the same sets of solutions  in the end, it is of interest to distinguish  the two different ways of looking at the solutions. 
In the construction of the reference solutions  some features of asymptotic simplicity are 
{\it built in from the start} by using asymptotic end data. In the stability result, however,  asymptotic simplicity for the perturbed solution is {\it deduced} as a consequence of the conformal properties of the equations and the reference solution.

In contrast to the approach of \cite{hadzic-speck-2015}, which concentrates on deriving suitable estimates, the emphasis is put in this article on the analysis of the field equations and  the explicit 
use of their conformal properties. 
While the conformal equations may lead to serious difficulties when the conformal structure of the solutions is intrinsically not well behaved at time-like infinity, they give results which are sharp and complete if the conformal structure extends smoothly and only the standard energy estimates for symmetric hyperbolic systems are needed.

Moreover, the information obtained on the equations is in that case of considerable practical interest. The reduced evolution system provides the possibility to calculate numerically  - on a finite grid - future complete solutions to Einstein's field equations, including the details of their asymptotic behaviour. In the Einstein-$\lambda$ case this has been successfully demonstrated by the work of Beyer (see \cite{beyer:2008} and the references given there).

Besides the one analysed in  \cite{friedrich:massive fields} this is the second example that illustrates that even in cases in which the energy momentum tensor is not trace free the  conformal field equations with $\lambda > 0$ and suitably rescaled matter fields can imply hyperbolic evolution equations that are well defined up to and beyond the future time-like infinity of the physical solutions. The two cases are quite different but the
results  suggest that the analysis of the asymptotic conformal structure in the presence of matter fields can be more useful than expected.

The possibility to extend  solutions to the conformal field equations into a domain in which $\Omega < 0$, where they define another solution to the original  equations (see section \ref{ex-strong-stab}), has been used here only as a technical device in the stability argument leading to 
Theorem \ref{main-result}. Whether it  is of any significance in the context of Penrose's proposal of {\it conformal cyclic cosmologies}
\cite{penrose:2011} is a question not discussed here.

\section{The Einstein-$\lambda$-dust system}
\label{Einstein-lambda-dust}

\vspace{.5cm}

The Einstein-Euler system with cosmological constant $\lambda$ consists of the Einstein equations 
\begin{equation}
\label{einst}
\hat{R}_{\mu \nu} - \frac{1}{2}\,\hat{R}\,\hat{g}_{\mu \nu} 
+ \lambda\,\hat{g}_{\mu \nu} = \kappa\,\hat{T}_{\mu \nu},
\end{equation}
for a Lorentz metric $\hat{g}_{\mu \nu}$ on a four-dimensional manifold $\hat{M}$ with an energy momentum tensor of a simple ideal fluid 
\begin{equation}
\label{spfl}
\hat{T}_{\mu \nu} = (\hat{\rho} + \hat{p})\,\hat{U}_{\mu}\,\hat{U}_{\nu} + 
\hat{p}\,\hat{g}_{\mu \nu}.
\end{equation}
Here $\hat{U}^{\mu}$ is the future directed time-like flow vector field,  normalized so that  
$\hat{U}_{\mu}\,\hat{U}^{\mu} = - 1$, and 
$\hat{\rho}$ and $\hat{p}$ denote the total energy density and the pressure as measured
by an observer moving with the fluid. The equations require the relation 
$\hat{\nabla}^{\mu}\,\hat{T}_{\mu \nu} = 0$,
which is equivalent to the system consisting of the equations 
\begin{equation}
\label{2pfleq}
(\hat{\rho} + \hat{p})\,\hat{U}^{\mu}\,\hat{\nabla}_{\mu}\,\hat{U}_{\nu}
+ \{\hat{U}_{\nu}\,\hat{U}^{\mu}\,\hat{\nabla}_{\mu} + \hat{\nabla}_{\nu}\}\,\hat{p} = 0, 
\end{equation}
\begin{equation}
\label{1pfleq}
\hat{U}^{\mu}\,\hat{\nabla}_{\mu}\,\hat{\rho} + (\hat{\rho} + \hat{p})\,\hat{\nabla}_{\mu}\,\hat{U}^{\mu} = 0.
\end{equation}
These equations must be implemented by an equation of state.

In the following we set $\kappa = 1$,  assume $\lambda > 0$, and consider solutions on manifolds diffeomorphic to $\hat{M} = \mathbb{R} \times S$ where $S$ is a compact (without boundary), orientable
$3$-manifold which specifies the topology of the time slices. We will be interested in the case where $\hat{p} = 0$ throughout, referred to as 
 {\it pressure free matter} or, shortly,  as {\it dust}. It is supposed  that $\hat{\rho}$ does not vanish identically and satisfies
\begin{equation}
\label{pos-en}
\hat{\rho} \ge  0 \quad \mbox{on} \quad \hat{M}.
\end{equation}
Equation (\ref{2pfleq}) reduces then to 
$\hat{\rho}\,\,\hat{U}^{\mu}\,\hat{\nabla}_{\mu}\,\hat{U}^{\nu} = 0$. 
This will be satisfied   without condition on $\hat{U}^{\mu}$ on sets where 
$\hat{\rho} = 0$ 
 and implies that the flow is geodesic where $\hat{\rho} \neq 0$. We require 
$\hat{U}^{\mu}$ to be geodesic everywhere.
The system to be considered consists then of  (\ref{einst}),
\begin{equation}
\label{dust}
\hat{T}_{\mu \nu} = \hat{\rho}\,\,\hat{U}_{\mu}\,\hat{U}_{\nu},
\end{equation}
\begin{equation}
\label{2dust}
\quad \quad \,
\hat{U}^{\mu}\,\hat{\nabla}_{\mu}\,\hat{U}^{\nu} = 0, \quad
\quad \hat{U}_{\mu}\,\hat{U}^{\mu} = - 1,
\end{equation}
\begin{equation}
\label{1dust}
\hat{\nabla}_{\mu}\,(\hat{\rho}\,\,\hat{U}^{\mu}) = 0. \quad \quad \quad \quad
\quad \,\,\,
\end{equation}

\vspace{.3cm}

Let $\hat{S}$ be a hypersurface in $\hat{M}$ which is space-like for $\hat{g}_{\mu\nu}$ and denote by $\hat{n}^{\mu}$ the future directed normal of $\hat{S}$ normalized by $\hat{n}_{\mu}\,\hat{n}^{\mu} = - 1$.  Let coordinates $x^{\mu}$ be given near $\hat{S}$ so that 
$\hat{S} = \{x^0 = 0\}$ and the $x^{\alpha}$, $\alpha, \beta  = 1, 2, 3$, are local coordinates on $\hat{S}$. Denote by 
$\hat{h}_{\alpha \beta}$, $\hat{\kappa}_{\alpha \beta}$
the first and the second fundamental form induced on $\hat{S}$ by $\hat{g}_{\mu \nu}$ and by 
$\hat{h}_{\mu}\,^{\nu} = \hat{g}_{\mu}\,^{\nu} + \hat{n}_{\mu}\,\hat{n}^{\nu}$ the orthogonal projector onto the tangent spaces of $\hat{S}$. Equations 
(\ref{2dust}), (\ref{1dust}) are evolution equations for $\hat{U}^{\mu}$ and $\hat{\rho}$.
Equation (\ref{einst})  induces with (\ref{dust}) on $\hat{S}$ the constraints
\[
0 = R[\hat{h}] - \hat{\kappa}_{\alpha \beta}\,\hat{\kappa}^{\alpha \beta} + (\hat{\kappa}_{\alpha}\,^{\alpha})^2 - 2\,\lambda - 2\,\hat{n}^{\mu}\,\hat{n}^{\nu}\,
\hat{T}_{\mu \nu},
\]
\[
0 = \hat{D}_{\beta}\,\hat{\kappa}_{\alpha}\,^{\beta} - \hat{D}_{\alpha}\,\hat{\kappa}_{\beta}\,^{\beta} 
- \hat{n}^{\mu}\,\hat{h}_{\alpha}\,^{\nu}\,\hat{T}_{\mu \nu}.
\]
Setting  $a = -  \hat{n}^{\mu}\,\hat{U}_{\mu} > 0$,  
$\hat{u}_{\mu} = \hat{h}_{\mu}\,^{\nu}\,\hat{U}_{\nu}$, so that
\[
\hat{U}_{\mu} = a\,\hat{n}_{\mu} + \hat{u}_{\mu} \quad \mbox{with} \,\,\, 
- 1 = - a^2 + \hat{u}_{\beta}\,\hat{u}^{\beta}
\quad \mbox{where} \quad
\hat{u}_{\beta}\,\hat{u}^{\beta} = \hat{h}^{\beta \gamma}\,\hat{u}_{\beta}\,\hat{u}_{\gamma},
\]
the constraints take the form
\begin{equation}
\label{hat-Ham-constr}
0 = R[\hat{h}] - \hat{\kappa}_{\alpha \beta}\,\hat{\kappa}^{\alpha \beta} 
+ (\hat{\kappa}_{\alpha}\,^{\alpha})^2 - 2\,\lambda 
- 2\,\hat{\rho}\,(1 +  \hat{u}_{\alpha}\,\hat{u}^{\alpha}),
\end{equation}
\begin{equation}
\label{hat-mom-constr}
0 = \hat{D}_{\beta}\,\hat{\kappa}_{\alpha}\,^{\beta} - \hat{D}_{\alpha}\,\hat{\kappa}_{\beta}\,^{\beta} 
+ \hat{\rho}\,\sqrt{1 +  \hat{u}_{\beta}\,\hat{u}^{\beta}}\,\,\hat{u}_{\alpha}.
\end{equation}

\vspace{.3cm} 

It has been shown in \cite{friedrich98} 
how to derive from equations  (\ref{einst}), (\ref{dust}), (\ref{2dust}), (\ref{1dust})
a symmetric hyperbolic evolution system of equations for all unknowns in a gauge based on the flow vector field $\hat{U}$. 
Given $\lambda > 0$ and a sufficiently smooth  initial data set 
\begin{equation}
\label{phys-in-data-set}
(\hat{S}, \,\hat{h}_{\alpha \beta}, \,\hat{\kappa}_{\alpha \beta}, \,\hat{u}^{\alpha}, \,\hat{\rho}), 
\end{equation}
satisfying (\ref{hat-Ham-constr}), 
(\ref{hat-mom-constr}) with  $\hat{h}_{\alpha \beta}$ a Riemannian metric and
$\hat{\rho} \ge 0$, the evolution  system
can be used to construct 
a globally hyperbolic solution $(\hat{M},\,\hat{g}_{\mu\nu},\,\hat{U}^{\mu}, \,\hat{\rho})$ 
to the Einstein-dust equations with cosmological constant $\lambda$ into which the initial data set is isometrically embedded so that $\hat{S}$ represents after an identification a space-like Cauchy hypersurface for 
$(\hat{M},\,\hat{g}_{\mu\nu})$. The manifold $\hat{M}$ will then be ruled by the geodesics tangent to $\hat{U}^{\mu}$. The ODE's
\[
\hat{U}^{\mu}\hat{\nabla}_{\mu}\,\hat{\rho} + \hat{\rho}\,\hat{\nabla}_{\mu}\,\hat{U}^{\mu} = 0,
\]
along the geodesics tangent to $\hat{U}^{\mu}$ 
ensure that $\hat{\rho} > 0$  or $= 0$ along a given geodesic, depending on whether this relation is satisfied at the point where the geodesic intersects $\hat{S}$. Thus 
$\hat{\rho} \ge 0$ will hold on $\hat{M}$.

For smooth initial data the evolution system given in  \cite{friedrich98} provides a smooth solution in coordinates $x^0 = t$, $x^a$ so that $<dx^a, \hat{U}> \,= 0$, $<dt, \hat{U}> \,= 1$, whence $\hat{U} = \partial_t$.  The initial hypersurfac  is given by $\hat{S} = \{t = t_*\}$ for some fixed value $t_*$,  the metric is of the form
\begin{equation}
\label{phys-coords}
\hat{g} = - (a\,dt)^2 + h_{\alpha \beta}\,(\hat{u}^{\alpha}\,dt +dx^{\alpha})\,
(\hat{u}^{\beta} \,dt+dx^{\beta})
\quad \mbox{on} \quad \hat{M},
\end{equation}
the future directed $\hat{g}$-unit normal to $\hat{S}$ is given by
\begin{equation}
\label{phys-unit-normal}
\hat{n}^{\mu} = \frac{1}{a}(\delta^{\mu} - \hat{u}^{\mu}) \quad \mbox{with shift vector field
$\hat{u}^{\mu}$ so that} \quad \hat{u}^0 = 0,
\end{equation}
and the lapse function $a$ satisfies $- 1 = \hat{g}(\hat{U}, \hat{U}) 
= - a^2 + h_{\alpha \beta}\,\hat{u}^{\alpha}\,\hat{u}{\beta}$.
If $\hat{U}$ is hypersurface orthogonal we can assume that $a = 1$, $\hat{u}^{\alpha} = 0$ and
the coordinates define a Gauss system. This will not necessarily be assumed in this article.

The questions to be analyzed in the following asks whether there exist a reasonably large set of data for which  the solutions can be extended to become future complete, so that $t$ takes values in 
$[t_*, \infty[$, and whether these solutions allow us to give a sharp and detailed description of the 
asymptotic behaviour of the conformal structure in the expanding direction, where
$t \rightarrow \infty$.


\section{The metric conformal field equations}
\label{conf-field-equ}

Let $\Omega$ denote a positive {\it conformal factor} on $\hat{M}$ and 
$g_{\mu \nu} = \Omega^2\,\hat{g}_{\mu\nu}$ the {\it rescaled metric}.
We shall in the following consider the tensor fields
\begin{equation}
\label{1-unknowns}
\Omega, \quad s = \frac{1}{4}\,\nabla_{\mu}\nabla^{\mu}\,\Omega + \frac{1}{24}\,\Omega\,R[g],
\quad 
L_{\mu\nu} = \frac{1}{2}\left(R_{\mu\nu}[g] - \frac{1}{6}\,R[g]\,g_{\mu\nu}\right),
\end{equation}
\begin{equation}
\label{2-unknowns}
W^{\mu}\,_{\eta \nu \lambda} = \Omega^{-1}\,C^{\mu}\,_{\eta \nu \lambda}[g], 
\end{equation}
where $\nabla_{\mu}$ denotes the Levi-Civita connection of $g$ and 
the last two fields denote the Schouten and the rescaled conformal Weyl tensor of $g_{\mu\nu}$ respectively. Moreover, we shall consider the {\it conformal matter  fields}
\[
U_{\mu} = \Omega\,\hat{U}_{\mu}, \quad \quad \rho = \Omega^{-3}\,\hat{\rho}.
\]
The vector fields $U^{\mu} = g^{\mu\nu}\,U_{\nu}$ and  
$\hat{U}^{\mu} = \hat{g}^{\mu\nu}\,\hat{U}_{\nu}$ are then related by 
\[
U^{\mu} = \Omega^{-1}\hat{U}^{\mu} \quad \mbox{so that} \quad  
g(U, U) = \hat{g}(\hat{U}, \hat{U}) = - 1.
\]
The tensor fields above satisfy the system of {\it conformal field equations}
(see \cite{friedrich:1991}, \cite{friedrich:massive fields})

\begin{equation}
\label{coord-alg-equ}
6\,\Omega\,s - 3\,\nabla_{\eta}\Omega\,\nabla^{\eta}\Omega - 
\lambda = - \frac{1}{4}\,\hat{T},
\end{equation}

\begin{equation}
\label{coord-Omega-equ}
\nabla_{\mu}\,\nabla_{\nu}\Omega + \,\Omega\,L_{\mu\nu} - s\,g_{\mu\nu}
= \frac{1}{2}\,\Omega\,T^*_{\mu\nu},
\end{equation}

\begin{equation}
\label{coord-s-equ}
\nabla_{\mu}\,s + \nabla^{\eta}\Omega\,L_{\eta\mu} 
= \frac{1}{2}\,\nabla^{\eta}\Omega\,T^*_{\eta \mu}
- \frac{1}{24\,\Omega}\,\nabla_{\mu}\,\hat{T},
\end{equation}

\begin{equation}
\label{coord-L-equ}
\nabla_{\nu}\,L_{\lambda \eta} 
- \nabla_{\lambda}\,L_{\nu \eta} - 
\nabla_{\mu}\Omega\,\,W^{\mu}\,_{\eta \nu \lambda} 
= 2\,\hat{\nabla}_{[\nu}\,\hat{L}_{\lambda] \eta},
\end{equation}

\begin{equation}
\label{coord-W-equ}
\nabla_{\mu}\,W^{\mu}\,_{\eta \nu \lambda} 
= 2\,\Omega^{-1}\,\hat{\nabla}_{[\nu}\,\hat{L}_{\lambda] \eta}.
\end{equation}
The right hand sides are determined by the trace 
\begin{equation}
\label{coord-T-trace}
\hat{T} = \hat{g}^{\eta \mu}\,\hat{T}_{\eta \mu} = - \hat{\rho} = - \Omega^3\,\rho,
\end{equation}
and the trace free part
\begin{equation}
\label{coord-T-trace-free-part}
T^*_{\eta \mu} =
 \hat{\rho}\left(\hat{U}_{\eta}\,\hat{U}_{\mu} + \frac{1}{4}\,\hat{g}_{\eta \mu}\right) = 
 \Omega\,\rho \left(U_{\eta}\,U_{\mu} + \frac{1}{4}\,g_{\eta \mu}\right),
\end{equation}
of the energy momentum tensor (\ref{dust}) and the  {\it physical Schouten tensor} $\hat{L}_{\mu \nu}$, which 
takes with our energy momentum tensor, the field equations, and the rescaled fields the form
\begin{equation}
\label{hat-L-form}
\hat{L}_{\mu \nu}  = 
\frac{1}{6}\,(\hat{\rho} + \lambda)\,\hat{g}_{\mu \nu}  + \frac{1}{2}\,\hat{\rho}\,\hat{U}_{\mu}\,\hat{U}_{\nu}
= \frac{1}{6}\,\lambda\,\hat{g}_{\mu \nu} 
+ \Omega\,\rho\left(\frac{1}{2}\,U_{\mu}\,U_{\nu} + \frac{1}{6}\,g_{\mu \nu}\right). 
\end{equation}

Taking into account the transformation law of the connection coefficients under conformal rescaling this gives
\[
2\,\hat{\nabla}_{[\nu} \hat{L}_{\lambda] \eta} 
= \hat{\nabla}_{[\nu} \hat{\rho} \,\,\hat{U}_{\lambda]} \,\hat{U}_{\eta} 
+ \frac{1}{3}\,\hat{\nabla}_{[\nu} \hat{\rho}\,\,\hat{g}_{\lambda] \eta}
+  \hat{\rho}\,(\hat{\nabla}_{[\nu}\hat{U}_{\lambda]} \,\hat{U}_{\eta} 
+  \hat{U}_{[\lambda} \,\hat{\nabla}_{\nu|}\hat{U}_{\eta})
\]
\[
= \Omega\left(\rho\,\,(\nabla_{[\nu}\,U_{\lambda]} \,U_{\eta} 
+  U_{[\lambda} \,\nabla_{\nu]}\,U_{\eta}) 
+ \nabla_{[\nu} \rho\,\,U_{\lambda]} \,U_{\eta} 
+ \frac{1}{3}\,\nabla_{[\nu} \rho \,\,g_{\lambda] \eta}\right)
\]
\[
+ \rho \left(\nabla_{[\nu} \Omega \,\,g_{\lambda] \eta}  +  2\,\nabla_{[\nu} \Omega\,\,U_{\lambda]} \,U_{\eta} 
+ U_{[\nu} \,g_{\lambda] \eta}\,g^{\pi \delta}\,\nabla_{\pi}\Omega\,U_{\delta}
\right).
\]
Finally, the geodesic equation (\ref{2dust})  translates into
\begin{equation}
\label{Omega-phys-geod}
\nabla_U U^{\mu} = \frac{1}{\Omega}\,(- g(U, U)\,g^{\mu}\,_{\rho} + U^{\mu}\,U_{\rho})\,\nabla^{\rho}\Omega.
\end{equation}
while equation (\ref{1dust}) for the density $\hat{\rho}$ gives 
\begin{equation}
\label{conformal-rho-equ}
\nabla_{U}\,\rho + \rho\,\nabla_{\mu}\,U^{\mu} = 0.
\end{equation}

\vspace{.2cm}

We express the equations in terms of a frame field
$e_k =e^{\mu}\,_k\partial_{x^{\mu}}$, $k = 0, 1, 2, 3$,  which has a time-like vector field given by 
\[
e_0 = U,
\]
and which  is orthonormal, so that $g_{jk} \equiv g(e_j, e_k) = \eta_{jk} = diag(-1, 1, 1, 1)$.
The space-like  frame fields are given by the $e_a$, where $a, b, c = 1, 2, 3$ denote spatial indices to which the summation convention applies. 
The metric is given by 
\[
g = \eta_{jk}\,\sigma^j\,\sigma^k,
\] 
where $\sigma^j$ denotes the field of 1-forms dual to $e_k$ so that their coefficients in the coordinates $x^{\mu}$ satisfy 
$\sigma^j\,_{\mu}\,e^{\mu}\,_k = \delta^j\,_k$.

The connection coefficients, defined by 
$\nabla_je_k \equiv \nabla_{e_j}e_k = \Gamma_j\,^l\,_k\,e_l$, 
satisfy $ \Gamma_{jlk} = - \Gamma_{jkl}$ with $ \Gamma_{jlk} = \Gamma_j\,^i\,_k\,g_{li}$,
because $\nabla_i g_{jk} = 0$. 
The covariant derivative of a tensor field $X^{\mu}\,_{\nu}$, given in the frame by 
$X^i\,_j$, takes the form
\[
\nabla_k\,X^i\,_j =  X^i\,_{j\,,\mu}\,e^{\mu}\,_k +  \Gamma_k\,^i\,_l \,X^l\,_j - 
\Gamma_k\,^i\,_l\,X^i\,_j.
\]
For the covariant version of $U$, i.e. 
$U_j = -  \,\delta^0\,_j$, equation (\ref{Omega-phys-geod}) implies the form
\begin{equation}
\label{nabla-U}
\nabla_k\,U_l = \Gamma_k\,^0\,_l 
= \delta^0\,_k\,\Omega^{-1}\,(\nabla_l\Omega + U_l\,\nabla_0\,\Omega)
+ \delta^a\,_k\,\delta^b\,_l\,\chi_{ab}.
\end{equation}
If $U$ is hypersurface orthogonal and if $\hat{S}$ were chosen to be orthogonal to $U$ so that the vector fields $e_a$ define an orthonormal frame on $\hat{S}$,
the field $\chi_{ab}$ would represent the second fundamental form induced by $g$ on the slice 
$\hat{S}$ whence  $\chi_{ab} = \chi_{(ab)}$. In general hypersurface orthogonality  will not be assumed here. 
We shall write $g^{ab}\,\chi_{ab} = \chi_a\,^a$.

\vspace{.1cm}

The metric coefficients and the connection coefficients satisfy the {\it first structural equations} 
\begin{equation}
\label{O-torsion-free condition}
e^{\mu}\,_{i,\,\nu}\,e^{\nu}\,_{j}
 - e^{\mu}\,_{j,\,\nu}\,e^{\nu}\,_{i} 
= (\Gamma_{j}\,^{k}\,_{i} - \Gamma_{i}\,^{k}\,_{j})\,e^{\mu}\,_{k},
\end{equation}
which ensures the connection to be torsion free, and  the {\it second structural equations}
\begin{equation}
\label{O-Ricci identity}
\Gamma_l\,^i\,_{j,\,\mu}\,e^{\mu}\,_k - \Gamma_k\,^i\,_{j,\,\mu}\,e^{\mu}\,_l
+ 2\,\Gamma_{[k}\,^{i\,p}\,\Gamma_{l]pj}
- 2\,\Gamma_{[k}\,^p\,_{l]}\,\Gamma_p\,^i\,_j
\end{equation}
\[
= \Omega\,W^i\,_{jkl}
+ 2\,\{g^i\,_{[k}\,L_{l] j} + L^i\,_{ [k}\,g_{l] j}\},
\]
which relates the coefficients (and thus the metric $g_{\mu\nu}$) to the unknowns in the conformal field equations. The conformal field equations read now

\begin{equation}
\label{f-alg-equ}
6\,\Omega\,s - 3\,\nabla_{i}\Omega\,\nabla^{i}\Omega - 
\lambda = \frac{1}{4}\,\Omega^3\,\rho,
\end{equation}

\begin{equation}
\label{f-Omega-equ}
\nabla_{j}\,\nabla_{k}\Omega + \,\Omega\,L_{j k} - s\,g_{j k}
= \frac{1}{2}\, 
 \Omega^2\,\rho\left(U_{j}\,U_{k} + \frac{1}{4}\,g_{j k}\right),
\end{equation}

\begin{equation}
\label{f-s-equ}
\nabla_{k}\,s + \nabla^{i}\Omega\,L_{ik} 
=  \frac{1}{2}\,\Omega\,\rho\,\nabla^{i}\Omega\left(U_{i}\,U_{k} + \frac{1}{4}\,g_{i k}\right)
+ \frac{1}{8}\,\Omega\,\rho\,\nabla_{k}\,\Omega + \frac{1}{24}\,\Omega^2\,\nabla_{k}\,\rho,
\end{equation}

\begin{equation}
\label{f-L-equ}
\nabla_{k}\,L_{l j} 
- \nabla_{l}\,L_{k j} - 
\nabla_{i}\Omega\,\,W^{i}\,_{j k l} 
\end{equation}
\[
= \Omega\left(\rho\,\,(\nabla_{[k}\,U_{l]} \,U_{j} 
+  U_{[l} \,\nabla_{k]}\,U_{j}) 
+ \nabla_{[k} \rho\,\,U_{l]} \,U_{j} 
+ \frac{1}{3}\,\nabla_{[k} \rho \,\,g_{l] j}\right)
\]
\[
+ \rho \left(\nabla_{[k} \Omega \,\,g_{l] j}  +  2\,\nabla_{[k} \Omega\,\,U_{l]} \,U_{j} 
+ U_{[k} \,g_{l] j}\,g^{p q}\,\nabla_{p}\Omega\,U_{q}
\right),
\]

\begin{equation}
\label{f-W-equ}
\nabla_{i}\,W^{i}\,_{j k l} 
= 
\end{equation}
\[
\rho\,\,(\nabla_{[k}\,U_{l]} \,U_{j} 
+  U_{[l} \,\nabla_{k]}\,U_{j}) 
+ \nabla_{[k} \rho\,\,U_{l]} \,U_{j} 
+ \frac{1}{3}\,\nabla_{[k} \rho \,\,g_{l] j}
+ \frac{1}{\Omega}\,\,\rho\,\,Z_{jkl}
\]
with 
\[
Z_{jkl} = 
\nabla_{[k} \Omega \,\,g_{l] j}  +  2\,\nabla_{[k} \Omega\,\,U_{l]} \,U_{j} 
+ U_{[k} \,g_{l] j}\,g^{p q}\,\nabla_{p}\Omega\,U_{q}.
\]
The matter equations are given by

\begin{equation}
\label{f-U-equ}
\nabla_U U^{k} = \frac{1}{\Omega}\,(g^{k}\,_{i} + U^{k}\,U_{i})\,\nabla^{i}\Omega,
\end{equation}
\begin{equation}
\label{f-rho-equ}
\nabla_{U}\,\rho + \rho\,\chi_a\,^a = 0.
\end{equation}

\vspace{.5cm}

Equations (\ref{O-torsion-free condition}) to (\ref{f-rho-equ}) establish a system of differential equations for the unknowns
\begin{equation}
\label{unknnowns}
e^{\mu}\,_k, \quad \Gamma_i\,^j\,_k, \quad \Omega, \quad s, \quad L_{jk}, \quad W^i\,_{jkl}, \quad
U^k, \quad \rho, 
\end{equation}
which is (apart from subtleties which may arise in cases  of low differentiability)  equivalent to the system
 (\ref{einst}), (\ref{dust}), (\ref{2dust}), (\ref{1dust}) in domains where $\Omega > 0$.

 If the system  is to be used to solve Cauchy problems with data given on a space-like hypersurface 
 $\hat{S}$, one has to restrict the 
available gauge freedom.  
We shall follow the procedure of \cite{friedrich:1991} and \cite{friedrich:massive fields}, where the conformal freedom is removed be considering the Ricci scalar $R = R[g]$ in a suitable neighborhood of $\hat{S}$ as a prescribed function of the space-time coordinates and by prescribing  suitable initial data for $\Omega$ and $\nabla_i\Omega$ on $\hat{S}$. The coordinates $\tau = x^0$ and $x^a$ are chosen near $\hat{S}$ so that 
$\tau = \tau_*$ on $\hat{S}$ and $<U, dx^a> \,= 0$, $\,\,<U, d\tau> \,= 1$, whence
\[
U^{\mu} = e^{\mu}\,_0 = \delta^{\mu}\,_0 \quad \mbox{ near $\hat{S}$}.
\]
Apart from a parameter transformation $t = t(\tau)$ these coordinates coincide with the
ones considered in (\ref{phys-coords}).
Precise conditions on the vector fields $e_a$ orthogonal to $U$ will be stated later.

\vspace{.3cm}

Our main interest is  the question whether there exist solutions  to the system above
on the domain where 
$\Omega > 0$ which admit a meaningful (i.e. sufficiently smooth) limit to a boundary where 
$\Omega \rightarrow 0$. In that case we write $\{\Omega = 0\} = {\cal J}^+$, and refer to this set as the future {\it conformal boundary} of the solution.
By equation (\ref{f-alg-equ}) the limit of $\nabla^i\,\Omega$ will then define a time-like normal to the set ${\cal J}^+$ so that the latter will define a space-like hypersurface. It represents (future) time-like and null infinity for the `physical'
space-time on which $\Omega > 0$.

There arises an obvious problem with the differential system above. 
The right hand sides of equations (\ref{f-W-equ}) and(\ref{f-U-equ}) are formally singular where
$\Omega \rightarrow 0$. This problem will be analyzed in the next section. Here we just point out its geometric nature.

If the fields entering equation (\ref{f-U-equ}) have limits as $\Omega \rightarrow 0$ the term in brackets on the right hand side of (\ref{f-U-equ}) defines a projection operator with kernel generated by the unit vector $U$. The right hand side of (\ref{f-U-equ}) can only admit a limit as $\Omega \rightarrow 0$ if the gradient of $\Omega$ is in the kernel of that operator and thus proportional to $U$, whence

\vspace{.1cm}

\noindent
\hspace*{.9cm}{\it The solutions can only admit a 
reasonably smooth {\it conformal boundary} \\ 
\hspace*{1.5cm}${\cal J}^+$ if the geodesics generated by $\hat{U}$ approach ${\cal J}^+$ orthogonally}.

\vspace{.1cm}

\noindent
Remarkably, the singularity of equation (\ref{f-W-equ}) is of a similar geometric nature. If we want to keep the freedom to have non-vanishing conformal densities $\rho$ on ${\cal J}^+$, the right hand side of  (\ref{f-W-equ})
can only assume a limit if $Z_{jkl} \rightarrow 0$ at ${\cal J}^+$. Since this implies that 
$U^j\,Z_{jkl}  = - \nabla_{[k} \Omega\,\,U_{l]} \rightarrow 0$, which implies in turn that  $Z_{jkl} \rightarrow 0$,
the conclusion above follows again.

\section{The regularizing relation}
\label{the-reg-rel}

A conformal geodesic in a given space-time $(\hat{M}, \hat{g})$ is a curve $x^{\mu}(\sigma)$ together with a 1-form field $b_{\nu}(\sigma)$ which satisfy the system of {\it conformal geodesic equations}
\[
\hat{\nabla}_{V} V^{\mu}
+ S(b)_{\lambda}\,^{\mu}\,_{\rho}\,V^{\lambda}\,V^{\rho} = 0, 
\]
\[
\hat{\nabla}_{V}b_{\nu} - \frac{1}{2}\, b_{\mu}\,S(b)_{\lambda}\,^{\mu}\,_{\nu}\,V^{\lambda} 
- \hat{L}_{\lambda \nu}\,V^{\lambda} = 0,
\] 
where $S(b)_{\lambda}\,^{\mu}\,_{\rho} = \delta_{\lambda}\,^{\mu}\,b_{\rho}
+ \delta_{\rho}\,^{\mu}\,b_{\lambda} - \hat{g}_{\lambda \rho}\,\hat{g}^{\mu \nu}\,b_{\nu}$
and $V^{\mu}(\sigma) =  \frac{dx^{\mu}}{d\sigma\,}$ denotes the tangent vector of the curve. Sometimes it will be convenient to write these equations  in the form
\begin{equation}
\label{a-hat(g)-conf-geod}
\hat{\nabla}_V V + 2<b, V>V - \hat{g}(V, V)\,b = 0,
\end{equation}
\begin{equation}
\label{b-hat(g)-conf-geod}
\hat{\nabla}_Vb\,  - <b, V>b + \frac{1}{2}\,\hat{g}(b, b)\,V - \hat{L}(V, \,.\,) = 0,
\end{equation}
where the index position should be clear from the above. 

For a conformal geodesic the initial data at a given point consist of its tangent vector and its 1-form at that point.
On a given space-time there exist thus more conformal geodesics than metric  geodesics. Moreover, there exists in general no particular relation between conformal and metric geodesics. The problem of interest here is, however, very special in this respect. 

\begin{lemma}
\label{geod-conf-geod-lemma}
Let $(\hat{M}, \hat{g})$ be a solution to the Einstein-dust system (\ref{einst}), (\ref{dust}), (\ref{2dust}),  (\ref{1dust}). Then the geodesics tangential to the vector field $\,\hat{U}$ coincide after a reparameterization
with the curves underlying certain conformal geodesics.
\end{lemma}

\noindent
{\bf Proof:} Suppose $\bar{x}^{\mu}(t)$ is a $\hat{g}$-geodesic with
$\frac{d\bar{x}^{\mu}}{dt\,} = \hat{U}^{\mu}(\bar{x}(t))$ and $(x^{\mu}(\sigma), b_{\nu}(\sigma))$
a conformal geodesics with $V^{\mu}(\sigma) =  \frac{dx^{\mu}}{d\sigma\,}$. Then there exists a parameter transformation $t = t(\sigma)$  so that $\frac{dt}{d\sigma} > 0$ and 
$x^{\mu}(\sigma) = \bar{x}^{\mu}(t(\sigma))$ if and only if
\begin{equation}
\label{V-hatU-rel}
V^{\mu}(\sigma) = \omega(\sigma)^{-1}\,\hat{U}^{\mu}(\bar{x}(t(\sigma))) \quad \mbox{with} \quad 
\omega^{-1} = \frac{dt}{d\sigma} > 0, \quad \quad   
\hat{g}(V, V) = - \omega^{-2}.
\end{equation}
For $x^{\mu}(\sigma)$ to be up to a reparametrization a geodesic we need to have  
a relation
\begin{equation}
\label{b-V-rel}
b_{\mu} = \alpha\,V_{\mu},
\end{equation}
with some function $\alpha = \alpha(\sigma)$ so that  (\ref{a-hat(g)-conf-geod})  reads 
\begin{equation}
\label{V-non-affine-geod}
\hat{\nabla}_{V} V^{\mu} + \alpha\,\hat{g}(V, V)\,V^{\mu} = 0.
\end{equation}
It follows then that
$2\,\omega^{-3}\,\hat{\nabla}_{V}\,\omega = \hat{\nabla}_{V}\,(\hat{g}(V, V)) = - 2\,\alpha\,\omega^{-4}$, whence
\begin{equation}
\label{alpha-omega-rel}
\alpha = - \omega\,\hat{\nabla}_V\,\omega.
\end{equation}
Basic for our result is that relations (\ref{hat-L-form}) and (\ref{V-hatU-rel}) give along $x^{\mu}(\sigma)$
\[
V^{\nu}\,\hat{L}_{\nu \mu} = \frac{1}{6}\,(\lambda - 2\,\hat{\rho})\,V_{\mu}, \quad \mbox{with}  \quad
\hat{\rho} = \hat{\rho}(\bar{x}^{\mu}(t(\sigma))).
\]
Inserting this and (\ref{b-V-rel}) into (\ref{b-hat(g)-conf-geod}) and observing (\ref{V-non-affine-geod}),
(\ref{alpha-omega-rel}) gives  the equation
\[
\omega\,\,\frac{d^2\omega}{d\,\sigma^2} - \frac{1}{2}\,\left(\frac{d\,\omega}{d \sigma}\right)^2
+ \frac{1}{6}\,(\lambda - 2\,\hat{\rho}(\bar{x}^{\mu}(t(\sigma)))) = 0,
\]
which provides with the relation 
\begin{equation}
\label{t-sigma-rel}
\frac{dt}{d\sigma} = \frac{1}{\omega},
\end{equation}
a system of ODE's  for $\omega = \omega(\sigma)$ and $t = t(\sigma)$ along 
$x^{\mu}(\sigma) = \bar{x}^{\mu}(t(\sigma)))$. Prescribing arbitrary initial data 
$t|_{\sigma_*} = t_*$, $\omega|_{\sigma_*}$, and 
$\frac{d\omega}{d\,\sigma}|_{\sigma_*}$ with $ \omega_* > 0$ at the point 
$x^{\mu}(\sigma_*) = \bar{x}^{\mu}(t_*))$
it can be solved. A straight forward calculation then shows  that 
\[
V^{\mu}(\sigma) = \frac{1}{\omega}\,\hat{U}^{\mu}(\bar{x}(t(\sigma)), 
\quad b_{\nu}(\sigma) = - \frac{d \omega}{d\sigma}\,\hat{U}^{\mu}(\bar{x}(t(\sigma)), 
\]
do indeed satisfy equations (\ref{a-hat(g)-conf-geod}) and (\ref{b-hat(g)-conf-geod}). $\Box$

\vspace{.1cm}

\noindent
It will later be important to note that the freedom to prescribe the initial data for $\omega$ gives the freedom to prescribe $\alpha$ arbitrarily at a given point.

\vspace{.1cm}

Conformal geodesics are of interest in the present context because  the curves underlying {\it conformal geodesics are  conformal invariants of a given conformal structure}:
If $g_{\mu \nu} = \Omega^2\,\hat{g}_{\mu \nu}$, where  $\Omega$ is a conformal factor as considered above
and $x^{\mu}(\sigma)$, $b_{\lambda}(\sigma)$ satisfy the conformal geodesic equations with respect to 
$\hat{g}_{\mu\nu}$,   
then
$x^{\mu}(\sigma)$, $f_{\nu}(\sigma)$ with 
\begin{equation}
\label{b-f-transf}
f_{\nu}(\sigma) = b_{\nu}(\sigma) - \Omega^{-1}\nabla_{\nu}\Omega|_{x(\sigma)},
\end{equation}
satisfy the conformal geodesics equations 
\begin{equation}
\label{1-g-f-conf-geod}
\nabla_V V + 2<f, V>V - g(V, V)\,f = 0,
\end{equation}
\begin{equation}
\label{2-g-f-conf-geod}
\nabla_Vf\,  - <f, V>f + \frac{1}{2}\,g(f, f)\,V - L(V, \,.\,) = 0,
\end{equation}
with respect to $g_{\mu\nu}$, where $\nabla$ and $L$ denote the Levi-Civita connection and the Schouten tensor of $g_{\mu \nu}$ 
(for this and further properties of conformal geodesics we refer to  \cite{friedrich:AdS}, \cite{friedrich:cg on vac}).
If $g(V, V) = - \theta^{-2}$ with 
$\theta > 0$ at a given point, equation (\ref{1-g-f-conf-geod}) gives
\[
\nabla_V \theta = \theta<V, f>,
\]
which shows that $\theta$ will stay positive and $x^{\mu}(\sigma)$ will be  time-like as long as $V$ and  $f$ remain sufficiently smooth. Equations  (\ref{1-g-f-conf-geod}), (\ref{2-g-f-conf-geod})
do not see the relation 
$g_{\mu \nu} = \Omega^2\,\hat{g}_{\mu \nu}$.
Thus, if $(\hat{M}, \hat{g})$ admits a smooth conformal boundary ${\cal J}^+$, one can arrange time-like conformal geodesics to extend smoothly to ${\cal J}^+$ with finite and non-vanishing tangent vector.

In the following we shall assume $V$ to be a conformal geodesic vector field which is related, as in 
(\ref{V-hatU-rel}), to the $\hat{g}$-geodesic vector field $\hat{U}$ by 
\begin{equation}
\label{B-V-hatU-rel}
V^{\mu}= \omega^{-1}\,\hat{U}^{\mu}.
\end{equation}
With the notation above we have then 
\[
\theta\,V^{\mu} = U^{\mu} = \Omega^{-1}\,\hat{U}^{\mu},
\]
and thus 
\begin{equation}
\label{theta-omega-Omega-rel}
\theta = \frac{\omega}{\Omega}, \quad \quad 
\nabla_U \theta = \theta<U, f>.
\end{equation}
Since $\theta$ stays smooth and positive if $U$ crosses the conformal boundary this has the remarkable 
consequence, used already in \cite{friedrich:AdS},
that $\omega$ goes to zero precisely where $\Omega$ does.

In terms of $U$ equation  (\ref{1-g-f-conf-geod}) takes the form
\begin{equation}
\label{first-g-conf-geod-equ}
\nabla_U U + <U,f>U - g(U, U)\,f = 0.
\end{equation}
Replacing in (\ref{2-g-f-conf-geod}) the field $V$ by $U = \theta\,V$ renders that equation
 in the form
\begin{equation}
\label{second-g-conf-geod-equ}
\nabla_Uf - <U,f>f +\frac{1}{2}\,g(f,f)U - L(U, \,.\,) = 0.
\end{equation}
This version of the conformal geodesic equations will be assumed from now on. The only effect of the transition is a reparametrization of 
$x^{\mu}(\sigma) \rightarrow x^{\mu}(\tau)$, $f_{\nu}(\sigma) \rightarrow f_{\nu}(\tau)$ where
$\sigma$ is replaced by a function $\sigma(\tau)$ so that 
\begin{equation}
\label{tau-sigma-rel}
\frac{d \tau}{d \sigma} = \frac{1}{\theta(x(\sigma))}. 
\end{equation}
In the following the parameter $\tau$ will be used.

With (\ref{theta-omega-Omega-rel}) and the relations obtained in the proof of 
Lemma \ref{geod-conf-geod-lemma} we get
\[
f_{\mu} = b_{\mu} - \Omega^{-1}\,\nabla_{\mu}\Omega = 
- \omega\,\nabla_V\omega\,\,\hat{g}_{\mu\nu}\,V^{\nu} - \Omega^{-1}\,\nabla_{\mu}\Omega
\]
\[
= 
- (\theta\,\Omega)\,\theta^{-1}\,\nabla_U(\theta\,\Omega)\,\Omega^{-2}\,g_{\mu\nu}\,\theta^{-1}\,U^{\nu} 
- \Omega^{-1}\,\nabla_{\mu}\Omega
\]
\[
= 
- (\theta^{-1}\,\nabla_U\theta 
+ \Omega^{-1}\,\nabla_U\Omega)\,U_{\mu} 
- \Omega^{-1}\,\nabla_{\mu}\Omega,
\]
\[
= 
- (<U, f> 
+ \Omega^{-1}\,\nabla_U\Omega)\,U_{\mu} 
- \Omega^{-1}\,\nabla_{\mu}\Omega,
\]
and thus the {\it regularising relation}
\begin{equation}
\label{f-Omega-U-rel}
\nabla_{\mu}\Omega = 
- (\nabla_U \Omega + \Omega<U, f>)U_{\mu}
- \Omega\,f_{\mu}. 
\end{equation}

This relation will play a critical role. It will be used  later  to obtain a hyperbolic system of evolution equations which extends in a regular way to the set $\{\Omega = 0\}$ and it will be used to set up a subsidiary system to show that constraints and gauge conditions are preserved  by the evolution system.
Here it is used to remove the singularities in equations (\ref{f-W-equ}) and (\ref{f-U-equ}). 
In fact, replacing in $Z_{jkl}$ the term $\nabla_k\Omega$  by the right hand side of  (\ref{f-Omega-U-rel}), we get (\ref{f-W-equ})  in the form 

\begin{equation}
\label{regular-div-W-equ}
\nabla_{i}\,W^{i}\,_{j k l}  = 
\nabla_{[k} \rho\,\,U_{l]} \,U_{j} 
+ \frac{1}{3}\,\nabla_{[k} \rho \,\,g_{l] j}
\end{equation}
\[
+ \rho\,\,(\nabla_{[k}\,U_{l]} \,U_{j} 
+  U_{[l} \,\nabla_{k]}\,U_{j}
- f_{[k}\,g_{l]j} - 2\,f_{[k}\,U_{l]}\,U_j - U_{[k}\,g_{l]j}\,U^i\,f_i).
\]

\vspace{.1cm}

\noindent
Using (\ref{f-Omega-U-rel}) to replace $\nabla_k\Omega$ on the right hand side of (\ref{f-U-equ}), the equation takes the form
\begin{equation}
\label{regular-f-U-equ}
\nabla_0 U^{k} + f^k  + U^k\,U_i\,f^i = 0,
\end{equation}
which is just (\ref{first-g-conf-geod-equ}) again. Equation (\ref{nabla-U}) is then replaced by the formally regular version
\begin{equation}
\label{reg-nabla-U}
\nabla_k\,U_l = \Gamma_k\,^0\,_l 
= (- \delta^0\,_k\,f_b + \delta^a\,_k\,\chi_{ab})\,\delta^b\,_l.
\end{equation}

\vspace{.2cm}

Finally we note that given sufficient asymptotic smoothness 
and an arrangement such that  $\Omega(x(\tau)) \rightarrow 0$ for some finite value of $\tau$, the relation 
\begin{equation}
\label{t-tau-rel}
\frac{dt}{d\tau} = \frac{1}{\Omega(x(\tau))},
\end{equation}
which follows from  (\ref{t-sigma-rel}), (\ref{theta-omega-Omega-rel}), (\ref{tau-sigma-rel})
implies with (\ref{coord-alg-equ}) that $t \rightarrow \infty$ as 
$\Omega(x(\tau)) \rightarrow 0$.

\section{The hyperbolic reduced equations}
\label{hyp-red-equ}

To extract from our equations a hyperbolic system we need to complete the gauge conditions for the 
$g$-orthonormal frame field $e_k$ satisfying  $e_0 = U$. The reduction procedure of the Einstein-dust system in \cite{friedrich98} employs a frame that is $\hat{g}$-parallely transported in the direction of  
$\hat{U}$. Since the field $U$ is not geodesic with respect to $g$ this cannot be done here.
We use instead  a frame whose vector fields  $X$ satisfy 
 the Fermi transport law
\[
0 = \mathbb{F}_UX \equiv \nabla_UX - g(X,  \nabla_U U)\,U + g(X, U)\, \nabla_U U,
\]
which has the properties: $\mathbb{F}_UU = 0$ and if $\mathbb{F}_UX = 0$,  $\mathbb{F}_UY = 0$ then $\nabla_U(g(X, Y)) = 0$.

On a given space-like hypersurface transverse to the flow line of $U$ 
we thus choose smooth fields $e_k$ with $e_0 = U$ such that  $g_{jk} = g(e_j, e_k) = \eta_{jk}$
and extend the $e_a$ away from the hypersurface by the requirement  that $\mathbb{F}_Ue_a = 0$. The smooth orthonormal frame field so obtained  is then closely related to the frame considered  in \cite{friedrich98}. In fact,
if $\hat{e}_k$ is a $\hat{g}$-orthonormal frame such that $\hat{e}_0 = \hat{U}$ and 
$\hat{\nabla}_{\hat{U}}\hat{e}_k = 0$, then $e_k = \Omega^{-1}\hat{e}_k$ is a 
$g_{\mu\nu} = \Omega^2\,\hat{g}_{\mu\nu}$-orthonormal frame with $e_0 = U$ and
$\mathbb{F}_Ue_a = 0$.

As a consequence of relation $\mathbb{F}_Ue_k = 0$ the connection coefficients satisfy 
\begin{equation}
\label{fermi-transp}
\Gamma_0\,^a\,_b = 0.
\end{equation}
The transport equation for the flow field $U$ is given by
(\ref{first-g-conf-geod-equ}). The coefficients  $U^{\mu} = e^{\mu}\,_0 = \delta^{\mu}\,_0$ have been  fixed by our choice of coordinates, however, and equation (\ref{regular-f-U-equ}) reduces to 
the relation
\begin{equation}
\label{U-acc}
\Gamma_0\,^a\,_0 = - f^a = - g^{ab}\,f_b
\quad \mbox{resp.} \quad \Gamma_0\,^0\,_a = - f_a,
\end{equation}
between the connection coefficients and the acceleration of $U$.
The remaining not necessarily vanishing connection coefficients are then given by
\begin{equation}
\label{U-acc}
\Gamma_a\,^b\,_c
\quad \mbox{and} \quad 
\Gamma_a\,^0\,_b = \nabla_a\,U_b = g(\nabla_{e_a}e_0, e_b) \equiv \chi_{ab}
\quad \mbox{resp.} \quad
\Gamma_a\,^b\,_0 =  \chi_a\,^b =  \chi_{ac}\,g ^{cb}.
\end{equation}
In the case in which $U$ resp. $\hat{U}$ is hypersurface orthogonal, the field $\chi_{ab}$ is symmetric and represents the second fundamental form while the  $ \Gamma_a\,^b\,_c$ are  the connection coefficients of the intrinsic connection induced on the hypersurfaces orthogonal to $U$  in the frame $e_a$.

We shall now derive the reduced equations for the remaining frame and connection coefficients.
With our gauge conditions and the connection coefficients above the first structural equations (\ref{O-torsion-free condition}) induce the evolution equations 
\begin{equation}
\label{frame-evolv}
e^{\mu}\,_{a, \,0}  = - f_a\,\delta^{\mu}\,_0 - \chi_a\,^b\,e^{\mu}\,_b,
\end{equation}
for the fields $e^{\mu}\,_a$.

 The second structural equations (\ref{O-Ricci identity}) induce the evolution equations
 \begin{equation}
 \label{space-Gamma-evol}
\Gamma_c\,^a\,_{b,\,0} = f^a\,\chi_{cb}
- \chi_{c}\,^a\,f_b
- \chi_{c}\,^d\,\Gamma_d\,^a\,_b
+ \Omega\,W^a\,_{b0c} - g^a\,_{c}\,L_{0 b} + L^a\,_{ 0}\,g_{c b},
\end{equation}
\begin{equation}
\label{chi-evol}
\chi_{a b,\,0} + D_a f_{b}
= f_{a}\,f_b
- \chi_{a}\,^c\,\chi_{cb}
- \Omega\,W_{0b0a}
+ L_{a b} - L_{ 00}\,g_{a b},
\end{equation}
for $\Gamma_c\,^a\,_b$ and $\chi_{a b}$, where we set 
\[
D_a f_b = f_{b\,,\mu}\,e^{\mu}\,_a - \Gamma_{a}\,^c\,_{b}\, f_{c}.
\]
No equation is implied  for $\Gamma_0\,^0\,_a = - f_a$ by  (\ref{O-Ricci identity}). 
Such an equation is provided, however,   by 
(\ref{second-g-conf-geod-equ}), which takes in our gauge the explicit form
\begin{equation}
\label{f0-evol}
f_{0,\,0} = - \frac{1}{2}\,f_j\,f ^j + L_{00},
\end{equation}
\begin{equation}
\label{fa-evol}
f_{a,\,0} = L_{0a}.
\end{equation}

At this stage arises a problem. We are aiming for a system that is symmetric hyperbolic. The principal part of the coupled 
system
\[
\chi_{a b,\,0} + D_a f_{b} = \ldots , \quad \quad \quad f_{a,\,0} = \ldots,
\]
does not satisfy the required symmetry condition. One might think of proceeding as follows.
The structural equations (\ref{O-Ricci identity}) imply after a contraction an 
analogue of Codacci's equation, which takes with the convention 
$D_c\,\chi_{a b} \equiv \chi_{a b\,,\mu}\,e^{\mu}\,_c - \Gamma_c\,^d\,_a\,\chi_{db}
- \Gamma_c\,^d\,_b\,\chi_{ad}$ the form
\[
D^a\,\chi_{a b} - D_b(\chi_a\,^a) = \ldots \,\,,
\]
(where the index position in the first term has to be respected because $\chi_{ab}$ is not necessarily symmetric). By adding a suitable multiple of this equation
to the second of the equations above one could hope to obtain a symmetric system. A careful analysis shows, however, that this does not work. We skip the details.

Help is again provided by (\ref{f-Omega-U-rel}). By this relation  the field 
\[
N_k =    \nabla_k\Omega  + (\nabla_U\Omega +
\Omega <U, f>)\,U_k +  \Omega\,f_k,
\]
vanishes in our gauge. While $N_0 = N_k\,U^k = 0$ identically,
the equation $N_a = 0$ with
\[
N_a = \Omega\,f_a + \nabla_a\Omega,
\]
has non-trivial content.  The relation 
\[
\nabla_j\,N_k =    \nabla_j\,\nabla_k\Omega  + 
\nabla_j\,(\nabla_U\Omega + \Omega <U, f>)\,U_k 
\]
\[
+ (\nabla_U\Omega + \Omega <U, f>)\,\nabla_j\,U_k 
+  \nabla_j\,\Omega\,f_k
+  \Omega\,\nabla_j\,f_k,
\]
implies in our gauge
\[
\nabla_a\,N_b - N_a\,f_b =    \nabla_a\,\nabla_b\,\Omega 
+ (\nabla_U\Omega + \Omega <U, f>)\,\chi_{ab}
- \Omega\,f_a\,f_b
+  \Omega\,\nabla_a f_b
\]
\[
=    \nabla_a\,\nabla_b\,\Omega 
+ (\nabla_U\Omega + \Omega <U, f>)\,\chi_{ab}
- \Omega\,f_a\,f_b
+  \Omega \,(D_a f_b - \chi_{a b}\,f_0) .
\]
\[
=    \nabla_a\,\nabla_b\,\Omega 
+ \nabla_U\Omega\,\chi_{ab}
- \Omega\,f_a\,f_b +  \Omega \,D_a f_b,
\]
which gives with (\ref{f-Omega-equ})
\begin{equation}
\label{N-and-nablaN=0}
\nabla_a\,N_b - N_a\,f_b 
=  \nabla_U\Omega\,\chi_{ab} + s\,g_{ab} +  \Omega \,(D_a\,f_b - f_a\,f_b
- L_{ab} + \frac{1}{8}\,\Omega\,\rho\,g_{ab}).
\end{equation}

Solving the equation $\nabla_a\,f_b - N_a\,f_b  = 0$ for $D_a f_b$ and using the resulting expression to replace that term in the evolution equation for $\chi_{ab}$, gives the latter in the form
\begin{equation}
\label{sing-chi-evol}
\chi_{a b,\,0} - \Omega^{-1}\,(\nabla_U\Omega\,\,\chi_{a b} + s\,g_{ab})
= 
- \chi_{a}\,^c\,\chi_{cb}
- \Omega\,W_{0a0b} - L_{ 00}\,g_{a b}.
\end{equation}
With the reduced equations obtained so far and the ones that follow below this gives again a symmetric hyperbolic system  where 
$\Omega \neq 0$. 

Let us assume that the solution admits a smooth conformal boundary ${\cal J}^+ = \{\Omega = 0\}$.
To obtain a system which extends in a regular fashion to ${\cal J}^+$ we recall that this would require that $e_0 = U$ approaches ${\cal J}^+$ orthogonally. With (\ref{f-alg-equ}) this would imply 
that
\[
\nabla_U\Omega \rightarrow - \nu < 0 \quad \mbox{as}\quad 
\Omega \rightarrow 0, \quad \mbox{where}\quad \nu \equiv \sqrt{- \frac{\lambda}{3}}, 
\]
and thus  $\nabla_U\Omega < 0$ also in a neighborhood of ${\cal J}^+$.
In the discussion of the conformal constraints on ${\cal J}^+$ in the next section we shall see that the conformal gauge can be chosen such that $s$ and $\chi_{ab}$ vanish at ${\cal J}^+$. If data on a `physical' initial hypersurface are evolved  in the direction of ${\cal J}^+$ it is, however, difficult to decide how the conformal gauge must be chosen such that these fields will vanish at ${\cal J}^+$. This suggests to introduce regularizing  unknowns which are derived from fields which go to zero at ${\cal J}^+$ in any conformal gauge. Such unknowns are suggested by the equation $\nabla_a\,f_b - N_a\,f_b  = 0$. In fact, 
the fields
\begin{equation}
\label{regularizing-unknowns}
\zeta_{ab} \equiv \frac{\chi_{a b} - \frac{1}{3}\,g_{ab}\,\chi_c\,^c}{\Omega}, \quad \quad 
\xi \equiv \frac{ \nabla_U\Omega\,\,\chi_c\,^c + 3\,s}{\Omega},
\end{equation}
satisfy for $\Omega \neq 0$ and $\nabla_U\Omega \neq 0$ by  (\ref{N-and-nablaN=0})  
\begin{equation}
\label{zeta-in-terms-of-f-etc}
\zeta_{ab} = 
 - (\nabla_U\,\Omega)^{-1} \left(D_a\,f_b - f_a\,f_b  - L_{ab} 
- \frac{1}{3}\,(D_c\,f^c - f_c\,f^c  - L_{c}\,^c)\,g_{ab}\right),
\end{equation}
and 
\begin{equation}
\label{xi-in-terms-of-f-etc}
\xi = - D_a\,f^a 
+ f_a\,f^a  + L_{a}\,^a - \frac{3}{8}\,\Omega\,\rho,
\end{equation}
and can thus be expected to extend smoothly to ${\cal J}^+$.
The original unknown will be recovered from the new ones by
\begin{equation}
\label{chi-in-terms-zeta-xi}
\chi_{a b}  = \Omega\,\zeta_{ab} +  \frac{1}{3}\, (\nabla_U\Omega)^{-1}\,(\Omega\,\xi - 3\,s)\,g_{ab}, 
\end{equation}
which will certainly be well defined on neighbourhoods of ${\cal J}^+$ where $\nabla_U\Omega \neq 0$. This will suffice for our purpose because we can use equation
(\ref{sing-chi-evol}) where $\Omega \neq 0$.

The equations we have obtained so far imply equations for the unknowns (\ref{regularizing-unknowns}) that are regular where $\nabla_U\Omega \neq 0$. Indeed, a direct calculation gives with 
(\ref{sing-chi-evol}) the equation
\begin{equation}
\label{zeta-evol}
\zeta_{ab\,,0} = - \Omega\,(\zeta_a\,^c\,\zeta_{cb} - \frac{1}{3}\,\zeta^{cd}\,\zeta_{dc}\,g_{ab})
- \frac{2}{3}\,(\nabla_U\Omega)^{-1}\,(\Omega\,\xi - 3\,s)\,\zeta_{ab} - W_{0a0b}.
\end{equation}
From (\ref{f-Omega-equ}) follows
\[
\Omega_{,00} - \Gamma_0\,^a\,_0\nabla_a\Omega = \nabla_0\nabla_0\Omega
= - \Omega\,L_{00} - s + \frac{3}{8}\,\Omega^2\,\rho,
\]
and thus with $0 = N_a = \Omega\,f_a + \nabla_a\Omega$ 
\[
\Omega_{,00} =  \Omega\,f_a\,f^a 
- \Omega\,L_{00} - s + \frac{3}{8}\,\Omega^2\,\rho.
\]
Equation (\ref{f-s-equ}) gives with  (\ref{f-rho-equ}) and $N_a = 0$
\[
s_{,0} = \nabla_U\Omega\,L_{00} + \Omega\,f^a\,L_{a0} - \frac{1}{4}\,\rho\,\Omega\,\nabla_U\Omega
- \frac{1}{24}\,\rho\,\Omega^2\,\chi.
\]
With these two equations relation  (\ref{sing-chi-evol}) implies 
\begin{equation}
\label{xi-evol}
\xi_{,0} = (\nabla_U\Omega)^{-1}\,(\Omega\,\xi - 3\,s) \left(- \frac{1}{3}\,\xi + f_a\,f^a - L_{00} + \frac{1}{4}\,\rho\,\Omega\right)
\end{equation}
\[
- \nabla_U\Omega\,\,\Omega\,\,\zeta_{cd}\,\zeta^{dc} 
+ 3\,f^a\,L_{a0} - \frac{3}{4}\,\rho\,\nabla_U\Omega.
\] 
This completes the evolution system for the metric and the connection coefficients.

To deal with equations of first order we introduce
\[
\Sigma_k = \nabla_k\Omega,
\]
as an unknown and use (\ref{f-Omega-equ}) to get the evolution equations
\begin{equation}
\label{Omega-evol} 
\nabla_0\Omega = \Sigma_0,
\end{equation}
\begin{equation}
\label{Sigma-evol}
\nabla_{0}\,\Sigma_{k} = -  \,\Omega\,L_{0 k} + s\,g_{0 k}
+ \frac{1}{2}\, \Omega^2\,\rho\left(U_{0}\,U_{k} + \frac{1}{4}\,g_{0 k}\right).
\end{equation}
From (\ref{f-s-equ}) we get
\begin{equation}
\label{s-evol}
\nabla_{0}\,s = -  \nabla^{i}\Omega\,L_{i0} 
=  \frac{1}{2}\,\Omega\,\rho\,\nabla^{i}\Omega\left(U_{i}\,U_{0} + \frac{1}{4}\,g_{i 0}\right)
+ \frac{1}{8}\,\Omega\,\rho\,\nabla_{0}\,\Omega + \frac{1}{24}\,\Omega^2\,\nabla_{0}\,\rho.
\end{equation}
As mentioned above, the Ricci scalar $R = R[g]$ of $g_{\mu\nu}$ will play the role of a conformal gauge source function  and thus be prescribed  as an explicit function of the coordinates near the initial hypersurface. 
Because of the relation
\begin{equation}
\label{L00-def}
 - L_{00} + g^{ab}\,L_{ab} = L_j\,^j =\frac{1}{6}\,R,
\end{equation}
it suffices to derive an evolution system for the components
$L_{0a}$, $L_{ab}$, $a, b = 1, 2, 3$, of the Schouten tensor. To simplify the equations  we set
\begin{equation}
\label{def-K}
K_{jkl} =
\nabla_{i}\Omega\,\,W^{i}\,_{j k l} 
\end{equation}
\[
 + \Omega\left(\rho\,\,(\nabla_{[k}\,U_{l]} \,U_{j} 
+  U_{[l} \,\nabla_{k]}\,U_{j}) 
+ \nabla_{[k} \rho\,\,U_{l]} \,U_{j} 
+ \frac{1}{3}\,\nabla_{[k} \rho \,\,g_{l] j}\right)
\]
\[
+ \rho \,(\nabla_{[k} \Omega \,\,g_{l] j}  +  2\,\nabla_{[k} \Omega\,\,U_{l]} \,U_{j} 
+ U_{[k} \,g_{l] j}\,g^{p q}\,\nabla_{p}\Omega\,U_{q}),
\]
so that (\ref{f-L-equ}) takes the form
\[
\nabla_{k}\,L_{l j} - \nabla_{l}\,L_{k j} = K_{jkl}.
\] 
It implies by contraction
\[
 \nabla_0\,L_{l 0}  - g^{bc}\,\nabla_b\,L_{l c}  = \frac{1}{6}\,\nabla_{l}\,R + K^j\,_{jl}.
\]
These equations are used to define the evolution system
\begin{equation}
\label{L0a-evol}
 \nabla_0\,L_{0a}  - h ^{bc}\,\nabla_b\,L_{a c} = \frac{1}{6}\,\nabla_{a}\,R + K^j\,_{ja}, \quad a = 1, 2, 3,
\end{equation}
\begin{equation}
\label{L11-etc-evol}
\nabla_{0}\,L_{a a} - \nabla_{a}\,L_{0 a} =  K_{a0a},\quad a = 1, 2, 3,
\end{equation}
\begin{equation}
\label{L12-etrc-evol}
2\,\nabla_{0}\,L_{a b} - \nabla_{a}\,L_{0 b} - \nabla_{b}\,L_{0 a}  =  K_{a0b} +  K_{b0a},
\quad a,  b  = 1, 2, 3, a \neq b.
\end{equation}
for the set of unknowns
\[
L_{01}, \quad L_{02}, \quad L_{03}, \quad L_{11}, \quad L_{12}, \quad 
L_{13}, \quad L_{22}, \quad L_{23}, \quad L_{33}.
\]
For given right hand sides the system will then be symmetric hyperbolic on a neighborhood of an initial hypersurface  on which $e^{\mu}_0 = \delta^{\mu}\,_0$ and on which $e^0\,_a$ is sufficiently small. Moreover, we find with our gauge conditions
\[
K^j\,_{ja} = - \frac{1}{2}\,\rho\,(\Omega\,f_a + \nabla_a\Omega),
\]
\[
K_{a0b} = \nabla_i\Omega\,W^i\,_{a0b} 
+ \frac{1}{2}\,\Omega \left(\rho\,\chi_{ba} + \frac{1}{3}\,\nabla_U\rho\,g_{ab} \right),
\]
and thus the important fact that on the right hand sides of the evolution system above 
only that derivative of $\rho$ occurs which can be removed by using the equation (\ref{f-rho-equ}), i.e.
\begin{equation}
\label{rho-evol}
\nabla_{U}\,\rho + \rho\,\chi_a\,^a = 0.
\end{equation}
This equation is assumed, of course, to be part of the  reduced system.

The following extraction of an evolution system for the rescaled conformal Weyl tensor from equation (\ref{regular-div-W-equ}) is close to the procedure  to obtain evolution equations for the conformal Weyl tensor discussed in \cite{friedrich98}, \cite{friedrich:rendall}, to which we refer for more details.
Let 
\[
h^j\,_k = g^j\,_k +U^j\,U_k, \quad \quad l^j\,_k  = g^j\,_k +2\,U^j\,U_k,
\]
denote the projection operator which maps the tangent spaces onto their subspaces $U^{\perp}$
orthogonal to $U$ and the reflection operator which maps $U$ onto $- \,U$ and induces the identity on   
 $U^{\perp}$ and consider the totally antisymmetric tensor densities
 \[
 \epsilon_{ijkl} =  \epsilon_{[ijkl]} \quad \mbox{with} \quad \epsilon_{0123} = 1
 \quad \mbox{and} \quad \epsilon_{jkl} = U^i\, \epsilon_{ijkl}.
 \]   
Further, define the $U$-electric part $w_{jl}$ and the $U$-magnetic part $w^*_{jl}$ of $W^{i}\,_{j k l}$ by setting 
\[
w_{jl} =   W_{i p k q}\,U^i \,h^p\,_j\,U^k\,h^q\,_l, \quad \quad 
w^*_{jl} =   \frac{1}{2}\,W_{i p m n}\,\epsilon^{mn}\,_{kq}\,U^i \,h^p\,_j\,U^k\,h^q\,_l, 
\]
so that these symmetric trace free  fields are given in our gauge essentially by their `spatial' components 
$w_{ab}$ and $w^*_{ab}$.

It will be convenient to write equation  (\ref{regular-div-W-equ}) in the form $F_{jkl}  = 0$ with
\begin{equation}
\label{Fjkl-def}
F_{jkl} = \nabla_{i}\,W^{i}\,_{j k l} 
- \nabla_{[k} \rho\,\,U_{l]} \,U_{j} 
- \frac{1}{3}\,\nabla_{[k} \rho \,\,g_{l] j}
\end{equation}
\[
- \rho\,\,(\nabla_{[k}\,U_{l]} \,U_{j} 
+  U_{[l} \,\nabla_{k]}\,U_{j}
- f_{[k}\,g_{l]j} - 2\,f_{[k}\,U_{l]}\,U_j - U_{[k}\,g_{l]j}\,U^i\,f_i).
\]
Inserting the  representation 
\[
W_{ijkl} = 2\,(l_{i[k}\,w_{l]j} - l_{j[k}\,w_{l]i} - U_{[k}\,w^*_{l] p}\,\epsilon^p\,_{ij}
- U_{[i}\,w^*_{j]p}\,\epsilon^p\,_{kl}),
\]
of the rescaled conformal Weyl tensor into the equations 
\begin{equation}
\label{P-ij-def}
0 = P_{ij} \equiv - F_{pkq}\,h^p\,_{(i}\,U^k\,h^q\,_{j)}   
+ \frac{1}{3}\,h_{ij}\,h^{kl}\,F_{pmq}\,h^p\,_k\,U^m\,h^q\,_l,
\end{equation}
\begin{equation}
\label{Q-ij-def}
0 = Q_{ij} \equiv - \frac{1}{2}\,F_{mpq}\,h^m\,_{(i}\,\epsilon_{j)}\,^{pq},
\end{equation}
the latter take the explicit form
\begin{equation}
\label{w-evol}
w_{ab,\,0} + D_c\,w^*_{d(b}\,\epsilon_{a)}\,^{cd} = \chi_{(a}\,^c\,w_{b)c} + 2\,\chi^c\,_{(a}\,w_{b)c}
- 2\,\chi_c\,^c\,w_{ab} 
\end {equation}
\[
 - h_{ab}\,\chi^{cd}\,w_{cd} - 2\,a_c\,w_{d(b}\,\epsilon_{b)}\,^{cd}
 - \frac{1}{6}\,\rho\,(3\,\chi_{(ab)} - h_{ab}\,\chi_c\,^c),
\]

\begin{equation}
\label{w*-evol}
w^*_{ab,\,0} - D_c\,w_{d(b}\,\epsilon_{a)}\,^{cd} =  \chi^c\,_{(a}\,w^*_{b)c}  - \chi_c\,^c\,w^*_{ab} 
\end {equation}
\[
+ 2\,a_c\,w_{d(a}\,\epsilon_{b)}\,^{cd} + \chi_{cd}\,w_{ef}\,\epsilon_{(i}\,^{ce}\,\epsilon_{j)}\,^{df},
\]
where we set, as before,
\[
 D_a\,w_{bc} = w_{bc, \,\mu}\,e ^{\mu}\,_a    -  \Gamma_a\,^d\,_b\,w_{dc}   -  \Gamma_a\,^d\,_c\,w_{bd}, 
 \]
etc. (The slight differences  with the analogues equations in \cite{friedrich98}, \cite{friedrich:rendall}
result from the use of the relation ${\cal L}_U\,w_{ij} = w_{ij,\,0} + 2\,\chi_{(i}\,^k   \,w_{k)j}$ for $w_{ab}$ and $w^*_{ab}$.) For given right hand side equations (\ref{w-evol}) and  (\ref{w*-evol}) represent a symmetric hyperbolic system for $w_{ab}$ and $w^*_{ab}$ if it is ignored that these fields are trace free.
Their trace-freeness will be taken care of by the construction of the initial data and then be preserved by the equations. Again it is important that no derivatives of the field $\rho$ occur on the right hand sides.

\vspace{.3cm}

If on the right hand sides the field $\nabla_k\Omega$ is replaced by $\Sigma_k$,
$\nabla_0\rho$ is removed by using (\ref{rho-evol}), 
$\chi_{ab}$ is replaced by (\ref{chi-in-terms-zeta-xi}), and $L_{00}$ is removed where it occurs (also in expressions like 
$\nabla_a\,L_{0b} = L_{0b,\,\mu}\,e ^{\mu}\,_a- \Gamma_a\,^k\,_0\,L_{kb} - \Gamma_a\,^k\,_b\,L_{0k}$)
by using (\ref{L00-def}), then equations (\ref{frame-evolv}),
(\ref{space-Gamma-evol}), (\ref{f0-evol}), (\ref{fa-evol}), (\ref{zeta-evol}),  (\ref{xi-evol}), (\ref{Omega-evol}), 
(\ref{Sigma-evol}), (\ref{s-evol}), (\ref{L0a-evol}), (\ref{L11-etc-evol}), (\ref{L12-etrc-evol}), (\ref{rho-evol}),
(\ref{w-evol}), (\ref{w*-evol}) represent, irrespectively of the sign of $\Omega$,  for suitably chosen initial data a quasi-linear symmetric hyperbolic evolution system for the unknowns 
\[
e^{\mu}\,_a, \quad \Gamma_c\,^a\,_{b}, \quad  f_k, \quad \zeta_{ab}, \quad \xi, \quad 
\Omega, \quad \Sigma_k, \quad s, \quad  L_{0a}, \quad L_{ab}, \quad  \rho, \quad
w_{ab}, \quad  w^*_{ab},
\]
where $\nabla_0\Omega \neq 0$. Where $\Omega \neq 0$ such an evolution system can be obtained by 
replacing $\zeta_{ab}$ and $\xi$ by $\chi_{ab}$ and using directly equation (\ref{sing-chi-evol}).
The characteristics of the systems so obtained are time-like or null with respect to the solution metric, i.e.
the metric 
$g_{\mu\nu}$ that satisfies $g_{\mu\nu}\,e ^{\mu}\,_j\,e ^{\nu}\,_k = \eta_{jk}$.

\section{Asymptotic end data}
\label{as-end-dat}

In section \ref{ex-strong-stab} we shall discuss the natural question how initial data for the reduced field equations are derived from solutions to the constraints  (\ref{hat-Ham-constr}, (\ref{hat-mom-constr}) induced by the Einstein-$\lambda$-dust system  on `physical' initial hypersurfaces.
The nature of the argument employed in \ref{ex-strong-stab} suggests, however, to consider first asymptotic data.

For solutions to Einstein's field equations with a positive cosmological constant which admit a smooth conformal boundary ${\cal J}^+$
it has been observed in the vacuum case \cite{friedrich:1986a}, in the case of  matter models involving conformally covariant matter models with $\hat{g}^{\mu\nu}\,\hat{T}_{\mu\nu} = 0$ \cite{friedrich:1991}, 
and also in the case of  a matter model with $\hat{g}^{\mu\nu}\,\hat{T}_{\mu\nu} \neq 0$
\cite{friedrich:massive fields} that the problem of providing initial data simplifies considerably if solutions to the constraints are constructed on that boundary. There is no need any longer to consider non-linear elliptic equations.
Assuming  that the solutions admit a smooth conformal boundary ${\cal J}^+ = \{\Omega = 0\}$, 
it will be shown in this section that 
the constraints induced on ${\cal J}^+$ by the conformal equations 
in the Einstein-dust case with a positive cosmological constant lead to  the same simplification. Moreover, in the particular case where $\rho > 0$ on ${\cal J}^+$ they simplify even further.   The solutions to the conformal Einstein-dust constraints can then in principle be constructed without solving any differential equation at all.

To construct the {\it asymptotic end data} on a 3-manifold which will later acquire the status of a smooth conformal boundary, let $S$ be a smooth, orientable, compact (though the latter is not really needed in the following discussion)
$3$-manifold. Assume that it represents a smooth conformal boundary ${\cal J}^+$ of an Einstein dust solution with cosmological constant  $\lambda > 0$. The conformal constraints induced on it must then be considered with an induced metric which is Riemannian and a conformal factor $\Omega$ which vanishes on $S$.
As seen earlier, the future directed conformal flow field $U$ must be  orthogonal to $S$. The conformal field equations will be considered in a frame $e_k$, $k = 0, 1, 2, 3$, on $S$  
so that $e_0 = U$ and the  $e_a$, $a = 1, 2, 3$, represent a frame on $S$ for the induced metric
\[
h_{ab} = g_{ab} = g(e_a, e_b) = diag(1, 1, 1),
\]
on $S$. The connection coefficients defined by $g$ in the frame $e_k$ are given again by $\nabla_k e _j = \Gamma_k\,^l\,_j\,e_l$.
As before  $h^j\,_k = g_j\,^k + U_j\,U^k$ denotes the orthogonal projector onto $S$.
By assumption we have  
$\Omega > 0$ in the past and $< 0$ in the future of $S$ and thus $e_0(\Omega) < 0$ on $S$.
Because $e_0$ is orthogonal to $S$ the field
\[
\chi_{ab} = \Gamma_a\,^0\,_b = g(\nabla_{e_a}e_0, e_b), 
\]
represents the second fundamental form induced on $S$ and is thus symmetric, while the 
$\Gamma_a\,^b\,_c$ define the connection coefficients on $S$ in the frame $e_a$ of the Levi-Civita connection $D$ defined by the intrinsic metric $h_{ab}$. 

The electric part $w_{jl} = W_{ipkq}\,U^i\,U^k\,h^p\,_j\,h^q\,_l$ of the rescaled conformal Weyl tensor is then represented by $w_{ab} = W_{0a0b}$ and $w^*_{ab} = \frac{1}{2}\,W_{0acd}\,\epsilon_b\,^{cd}$
represents its magnetic part
$w^*_{jl} = \frac{1}{2}\,W_{ipmn}\,\epsilon^{mn}\,_{kq}\,U^i\,U^k   \,h^p\,_j\,h^q\,_l$, where 
$\epsilon_{ijkl}$ and $\epsilon_{jkl}$ are defined as before.

With these assumptions equation (\ref{f-alg-equ}) reduces to the condition
\begin{equation}
\label{dOmega-on-S}
\nabla_0\Omega  = - \nu,
 \quad \nabla^0\Omega = \nu 
\quad \mbox{on} \quad S, \quad \mbox{where} \quad \nu = \sqrt{\lambda/3} > 0.
\end{equation}
Equation (\ref{f-Omega-equ}) reduces on $S$ to $\nabla_{i}\,\nabla_{j}\Omega = s\,g_{ij}$.
The only non-trivial condition  implied by this relation  is
a restriction on the second fundamental form
\begin{equation}
\label{2nd-fund-form-on S}
\nu\,\chi_{a b } = s\,h_{ab} 
\quad \mbox{on} \quad S.
\end{equation}
Equation (\ref{f-s-equ}) implies the constraint
\begin{equation}
\label{L0a-on-S}
\nabla_a\,s + \nu\,L_{0 a}
= 0 \quad \mbox{on} \quad S.
\end{equation}
Under the conformal gauge transformation $g \rightarrow \bar{g} = \theta^2\,g$, $\Omega \rightarrow \bar{\Omega} = \theta\,\Omega$ with smooth $\theta > 0$ the function $s$ transforms as
$ s \rightarrow \bar{s} = \theta\,s + g^{\rho \delta}\,\nabla_{\rho}\Omega\,\nabla_{\nu}\theta$.
This shows that for given $\theta > 0$ on $S$ the derivative $\nabla_{\mu}\theta$ 
can be determined on $S$ such that $\bar{s}$ coincides on $S$ with any prescribed function. 
The function $s$ could be carried along as a free function in the following equations but for simplicity the choice 
that 
\begin{equation}
\label{s=0-etc}
s = 0, \quad \chi_{a b } = 0, \quad  \nabla_{i}\,\nabla_{j}\Omega = 0, \quad  L_{0a} = L_{a0} = 0 \quad \mbox{on} \quad S,
\end{equation}
will be assumed,  which still leaves the freedom to rescale the metric on $S$. It should be observed, however, that the gauge above may not be satisfied if a solution is evolved  into $S$ from the domain where $\Omega > 0$.
In that case the more general relations like (\ref{2nd-fund-form-on S}) and (\ref{L0a-on-S}) must be considered.

Because  the conformal Weyl tensor $\Omega\,W^i\,_{jkl}$ vanishes on $S$, the curvature tensor of $g$  is determined there by its Schouten tensor $L_{jk}$. Because the second fundamental form vanishes on $S$, the orthogonal projection of the curvature tensor of $g$ onto $S$ coincides by Gauss' theorem with the curvature tensor of $h$, i.e. $R_{abcd}[g] = R_{abcd}[h]$. It follows that the decomposition 
of $R_{abcd}[g]$ in terms $g_{ab} = h_{ab}$ and the components $L_{ab}[g]$ of its Schouten tensor
is formally identically with the decomposition of  of $R_{abcd}[h]$ in terms $h_{ab}$ and its Schouten tensor  $l_{ab}[h] = R_{ab}[h] - \frac{1}{4}\,R[h]\,h_{ab}$. This implies that 
\[
L_{ab}[g] = l_{ab}[h],
\] 
which can be calculated from $h_{ab}$. The component $L_{00}$ then follows from $\frac{1}{6}\,R[g] = L_j\,^j$ as
\[
L_{00} = - \frac{1}{6}\,R[g] + h^{ab}\,L_{ab},
\]
once the conformal gauge source function $R[g]$ has been prescribed.

Equation (\ref{f-L-equ}) induces the constraint
$\nabla_{a}\,L_{b c} - \nabla_{b}\,L_{a c} = - \nu\,W^{0}\,_{c ab}$ on $S$.
Because the second fundamental form on $S$ vanishes, it can be written in the form
\begin{equation}
\label{Cotton-magn-part}
w^*_{ab}  = \frac{1}{\nu}\,\epsilon_a\,^{cd}\,D_c\,l_{db}.
\end{equation}
The equation says that the magnetic part of the rescaled conformal Weyl tensor
is given on  $S$ up to a factor by the (dualized) Cotton tensor of $h$.
Equation (\ref{f-L-equ}) induces the further constraint
$\nabla_{a}\,L_{b 0} - \nabla_{b}\,L_{a 0} = 0$ on $S$. This is satisfied as a consequence of
(\ref{s=0-etc}).

\vspace{.2cm}

With $F_{jkl}$ given by (\ref{Fjkl-def}), the constraints induced on $S$ by equation 
(\ref{regular-div-W-equ}) are given by (see \cite{friedrich98}, \cite{friedrich:rendall})
\begin{equation}
\label{P-k-and-Q-k-def}
0 = P_k \equiv F_{jpl}\,U^j\,h^p\,_k\,U^l, \quad \quad  
0 = Q_k \equiv - \frac{1}{2}\,F_{jpq}\,U^j\,\epsilon_k\,^{pq}. 
\end{equation}
They can be written more explicitly in the form
\begin{equation}
\label{asymp-el-part-constr}
D^aw_{ac} = \frac{1}{3}\,D_{c} \rho - \rho\,f_c,
\end{equation}
which is a genuine constraint, and 
\begin{equation}
\label{asymp-mag-part-constr}
D^a\,w^*_{ab} = 0,
\end{equation}
which is, consistent with (\ref{Cotton-magn-part}), the differential identity satisfied by the Cotton tensor 
and imposes thus no additional restriction.

\vspace{.1cm}

The $1$-form $f_a$ characterizes the deviation of $U$ from hypersurface orthogonality
(see the datum $\hat{u}^{\alpha}$ in (\ref{phys-in-data-set})
and the following discussion of hypersurface orthogonal flows) and can be prescribed freely on $S$. The value of $f_0$ only affects the gauge. It can be prescribed freely and we assume that $f_0 = 0$ on $S$.

\vspace{.1cm}

The initial data for $\zeta_{ab}$ and $\xi$ which follow from
(\ref{zeta-in-terms-of-f-etc}) and (\ref{xi-in-terms-of-f-etc}) are then given on  $S$ by
\begin{equation}
\label{c-zeta-in-terms-of-f-etc}
\zeta_{ab} = 
\nu^{-1} \left(D_a\,f_b - f_a\,f_b  - L_{ab} 
- \frac{1}{3}\,(D_c\,f^c - f_c\,f^c  - L_{c}\,^c)\,g_{ab}\right),
\end{equation}
and 
\begin{equation}
\label{c-xi-in-terms-of-f-etc}
\xi = - D_a\,f^a 
+ f_a\,f^a  + L_{a}\,^a.
\end{equation}

\vspace{.1cm}

The observations above can be summarized in terms of local coordinates $x^{\alpha}$, $\alpha = 1, 2, 3$, on $S$ as follows.

\begin{lemma}
\label{free-data-on-scri}
Any smooth  initial data set for the reduced equations is determined on the set $S = \{\Omega = 0\}$
uniquely by a Riemannian metric $h_{\alpha \beta}$, the density $\rho \ge 0$, the acceleration 
$f_{\alpha}$ and a symmetric, $h$-trace free tensor field $w_{\alpha \beta}$, which are arbitrary up to the relation
\begin{equation}
\label{w-constr-on-scri}
D^{\alpha}w_{\alpha \beta} = \frac{1}{3}\,D_{\beta} \rho -  \rho\,f_{\beta} \quad \mbox{on} \quad S,
\end{equation}
where $D$ denotes the Levi-Civita operator defined by $h_{\alpha \beta}$.
\end{lemma}

As in the cases mentioned in the beginning there is no need to solve an analogue of the  Hamiltonian constraint.
The Riemannian space $(S, h_{\alpha \beta})$ is not subject to any further restriction. The situation even simplifies for
the class of data with $\rho > 0$ on $S$. In that case $h_{\alpha \beta}$, $\rho > 0$, and $w_{\alpha \beta}$ can be prescribed completely freely and $f_{\beta}$ is then determined by reading  (\ref{w-constr-on-scri}) as its defining equation. It should be pointed out, however, that if 
$f_{\alpha}$ is required to satisfy some extra conditions,  as  in the hypersurface orthogonal case discussed below, equation (\ref{w-constr-on-scri}) must be read as a differential equation. The situation can  then  be discussed by the well known splitting techniques 
used in the discussion of the standard constraints \cite{bartnik:isenberg}.  

\vspace{.1cm}

The gauge requirement  $s|_{\{\Omega = 0\}} = 0$ leaves the conformal gauge freedom 
\[
\Omega \rightarrow \Omega' = \theta\,\Omega, \quad 
g_{\mu\nu} \rightarrow g'_{\mu\nu} = \theta^2\,g_{\mu\nu}, 
\]
with smooth functions $\theta > 0$ that are arbitrary on $S$. 
If $n^{\mu}$ denotes  the future directed unit normal  to $S$
the conformal gauge transformation above  implies associated transformations
\[
\quad 
h_{\alpha \beta} \rightarrow h'_{\alpha \beta} = \theta^2\,g_{\alpha \beta}, \quad 
n^{\mu} \rightarrow n'^{\mu} = \theta^{-1}\,n^{\mu}, \quad 
U^{\mu} \rightarrow U'^{\mu} = \theta^{-1}\,U^{\mu}, \quad 
\rho \rightarrow \rho' = \theta^{-3}\,\rho,
\]
and, by the transformation law for the 1-forms associated with conformal geodesics,
\begin{equation}
\label{conf-transf-fa-on-scri}
f_{\alpha} \rightarrow f'_{\alpha} = f_{\alpha} - \theta^{-1}\,D_{\alpha}\theta.
\end{equation}
If $n$ is extended as unit vector field into $\hat{M}$, the relation 
$g_{\alpha \mu}\,W^{\mu}\,_{\nu \beta \rho}\,n ^{\nu}\,n^{\rho} = \Omega^{-1}\,g_{\alpha \mu}\,C^{\mu}\,_{\nu \beta \rho}\,n ^{\nu}\,n^{\rho}$ makes sense and suggests  on $S$  for $w_{\alpha \beta}$
the transformation law
\[
w_{\alpha \beta} \rightarrow w'_{\alpha \beta} = \theta^{-1}\,w_{\alpha \beta}.
\]
It follows then 
\[
h'^{\alpha \beta}\,D'_{\alpha}\,w'_{\rho \gamma} 
= \theta^{-3}\,h^{\alpha \beta}\,D_{\alpha}\,w_{\rho \gamma},
\] 
whence 
\[
D'^{\alpha}\,w'_{\alpha \beta} - \frac{1}{3}\,D'_{\beta}\,\rho' + \rho'\,f'_{\beta}
= \theta^{-3}\,(D_{\alpha}\,w^{\alpha}\,_{\beta} - \frac{1}{3}\,D_{\beta}\,\rho + \rho\,f_{\beta}),
\]
so that 
the constraints are preserved.

\subsection{Hypersurface orthogonal flows}
\label{hyp-orthog-flow}

Obviously, the vector field  $\hat{U}^{\mu}$ is   hypersurface orthogonal  where $\Omega \neq 0$  if and only if this is true for  $U^{\mu} = \Omega^{-1}\,\hat{U}^{\mu}$. Formally this follows from the relation 
$\hat{U}_{[\rho}\,\hat{\nabla}_{\mu}\,\hat{U}_{\nu]} = \Omega^{-2}\,U_{[\rho}\,\nabla_{\mu}\,U_{\nu]}$.
In our gauge the hypersurface orthogonality condition 
$\hat{U}_{[\rho}\,\hat{\nabla}_{\mu}\,\hat{U}_{\nu]} = 0$ is equivalent to  
\begin{equation}
\label{conf-hyp-ortho-cond}
0 = \nabla_{[a}\,U_{b]} = \chi_{[ab]}.
\end{equation}
From (\ref{chi-evol}) we get with $\sigma_{ab} = \chi_{(ab)}$ along the flow lines of $U^{\mu}$  the ODE
\[
\chi_{[a b],\,0} + D_{[a}\,f_{b]}
= \sigma_{a}\,^c\, \chi_{[cb]} - \sigma_{b}\,^{c}\, \chi_{[ca]}.
\]
It follows that $D_{[a}\,f_{b]} = 0$ if $U^{\mu}$ is hypersurface orthogonal.
If the solution admited a smooth conformal extension, so that $\chi_{[ab]} = 0$ on ${\cal J}^+$, 
we could conclude from the equation above that $\chi_{[ab]} = 0$ if we knew that $D_{[a}\,f_{b]} = 0$. 
With the gauge condition  $\nabla_a\,N_b - N_a\,f_b = 0$ equation (\ref{N-and-nablaN=0}) gives, however, only the relation
\[
0 = \Omega_{,\,0}\,\chi_{[ab]} +  \Omega \,D_{[a}\,f_{b]}. 
\]
But this combines with the equation above to  give
\[
\left(\Omega^{-1}\,\chi_{[a b]}\right)_{,0} = 
\sigma_{a}\,^c\,\left(\Omega^{-1}\,\chi_{[c b]}\right)
- \sigma_{b}\,^{c}\, \left(\Omega^{-1}\,\chi_{[ca]}\right).
\]
It follows that $\chi_{[a b]} = 0$ along a given integral curve of $U^{\mu}$ if it vanishes at a point of it where $\Omega \neq  0$. 
On the other hand, the relation above shows that $\Omega^{-1}\,\chi_{[a b]}$
assumes the limit $(\nabla_0\Omega)^{-1}\,D_{[a}\,f_{b]}$ on ${\cal J}^+$, which vanishes where the integral curves of $U^{\mu}$ meet ${\cal J}^+$ if and only if 
$D_{[a}\,f_{b]} = 0$ there. Observing the discussion of the conformal gauge freedom in the construction of data on the conformal boundary, in particular (\ref{conf-transf-fa-on-scri}), we conclude:

\begin{lemma}
\label{U-hyp-orthog-cond} 
Let be given a solution to the Einstein-dust system (\ref{einst}), (\ref{dust}), (\ref{2dust}),  (\ref{1dust})
that admits a smooth conformal boundary ${\cal J}^+$.
Then the  field $U^{\mu}$ is hypersurface orthogonal if and only if the initial data for the conformal field equations  induced on ${\cal J}^+$ in the gauge above  are such that  
\[
D_{[a}\,f_{b]} = 0 \quad \mbox{on} \quad {\cal J}^+.
\]
If this condition is satisfied and the field $f_a$ can be given on ${\cal J}^+$ as the differential of a function 
$f$, then the conformal gauge can be chosen so that $f_a = 0$ on ${\cal J}^+$.

\end{lemma}

\subsection{FLRW-type solutions}
\label{FLRW-sols}

\vspace{.5cm}

In the following we  discuss the FLRW solutions along the lines of the previous sections. 
The FLRW-type solutions to (\ref{einst}), (\ref{dust}), (\ref{2dust}), (\ref{1dust}) on 
$\hat{M} = \mathbb{R} \times S$ with $S = \mathbb{S}^3, \, \mathbb{T}^3$ or $\mathbb{H}^3_*$ 
(a suitable factor space of hyperbolic $3$-space) 
 are of the form
\[
\hat{g} = - dt^2 +  a^2\,k, \quad \quad \hat{U} = \partial_{t}, 
\quad \quad \hat{\rho} = \hat{\rho}(t) \ge 0,
\]
with a function $a = a(t) > 0$ and a $3$-metric  of constant curvature which is given in  local coordinates 
$x^{\alpha}$, $\alpha, \beta, \ldots = 1, 2, 3$, on $S$ by
$k =  k_{\alpha \beta}\,dx^{\alpha}\,dx^{\beta} $, so that $R_{\alpha \beta \gamma \delta }[k] = 2\,\epsilon\,k_{\alpha [\gamma}\,k_{\beta]\delta} $ where 
$\epsilon = 1, 0$ or $-1$. Rescaling the fields with a conformal factor $\Omega = \Omega(t)$ 
\[
\hat{g} \rightarrow g = \Omega^2\,\hat{g}, \quad \quad 
\hat{U} \rightarrow U =   \Omega^{-1}\,\hat{U} , \quad \quad 
\hat{\rho} \rightarrow \rho =  \Omega^{-3}\,\hat{\rho},
\]
and introducing  a coordinate  $x^0 = \tau(t)$ so that $<U, d\tau> \,\,= 1$, the conformal version of the metric above takes the form
\[
g = - d\tau^2 +  l^2\,k, \quad \quad U = \partial_{\tau}, 
 \quad \rho = \rho(\tau),
\]
with some function $l = l(\tau) > 0$. 
The non-vanishing Christoffel symbols  and the second fundamental form $\chi_{\alpha \beta}$ of the slices $\{\tau = const.\}$ are then given by
\[
\chi_{\alpha \beta} = \Gamma_{\alpha}\,^0\,_{\beta}[g] = l\,l'\,k_{\alpha \beta},  \quad
\Gamma_0\,^{\alpha}\,_{\gamma}[g] = \Gamma_{\gamma}\,^{\alpha}\,_0[g] 
= \frac{1}{l}\,l'\,k^{\alpha}\,_{\gamma}, \quad
\Gamma_{\beta}\,^{\alpha}\,_{\gamma}[g] =  \Gamma_{\beta}\,^{\alpha}\,_{\gamma}[k], 
\]
where $' = \frac{d}{d\tau}$. The Ricci scalar and the Schouten tensor are given by
\[
R[g] = \frac{6}{l^2}\,(\epsilon + l\,l'' + (l')^2), 
\]
\[
L_{00}[g] = \frac{1}{2\,l^2}\,(\epsilon - 2\,l\,l'' + (l')^2), \quad 
L_{\alpha 0}[g] = L_{0 \alpha}[g] = 0, \quad 
L_{\alpha \beta}[g] = \frac{1}{2}\,(\epsilon + (l')^2)\,k_{\alpha \beta}.
\]
Choosing  the conformal  gauge function as  $R[g] = 6\,\epsilon$ on $\hat{M}$, the function $l$ must satisfy $l\,l'' + (l')^2 + \epsilon\,(1 - l^2) = 0$. Using the remaining conformal gauge freedom to achieve $l =1$, $l' =0$ on a slice $\{\tau = const.\}$, it follows that $l = 1$.
The only non-vanishing Christoffel symboly are then given by 
$\Gamma_{\beta}\,^{\alpha}\,_{\gamma}[g] =  \Gamma_{\beta}\,^{\alpha}\,_{\gamma}[k]$ and
\[
L_{00} = \frac{\epsilon}{2}, \quad 
L_{\alpha0} = L_{0\alpha} = 0, \quad 
L_{\alpha \beta} = \frac{\epsilon}{2}\,k_{\alpha \beta}.
\]
Where $\Omega > 0$ the physical field is then given by
\begin{equation}
\label{hom-g-hat}
\hat{g} = \Omega^{-2}\,g = - dt^2 + a^2\,d\omega^2,
\end{equation}
\begin{equation}
\label{c-coord-conf-transf}
a(t) = \frac{1}{\Omega(\tau(t))}, \quad \quad 
\frac{dt}{d\tau} = \frac{1}{\Omega(\tau)}.
\end{equation}

\vspace{.1cm}

The high symmetry assumptions leads to a simplification of the conformal field equations.
There do not occur singularities any longer  in the equations.  
In fact, because $U$ is $g$-geodesic and hypersurface orthogonal and $\Omega = \Omega(\tau)$, the singularity in (\ref{Omega-phys-geod}) is gone.
Because the line element $g$ is locally conformally flat it follows that $W^{\mu}\,_{\nu \rho \kappa}  = 0$ and thus $\hat{\nabla}_{[\nu}\,\hat{L}_{\lambda] \rho} = 0$ 
by (\ref{coord-W-equ}).
Moreover,  it follows by (\ref{coord-L-equ}) that $\nabla_{[\nu}\,L_{\lambda] \rho} = 0$.

It will be assumed in the following that the conformal time coordinate $\tau$ vanishes on a set
$\{\Omega = 0\}$ and that $\nabla_U \Omega = \Omega' < 0$ there. Equations (\ref{coord-alg-equ}) and 
(\ref{coord-T-trace}) then  imply 
\[
\Omega'(0)  = - \nu = - \sqrt{\lambda/3} < 0.
\]
Equation (\ref{conformal-rho-equ}) 
reduces because of $\nabla_{\mu}\,U^{\mu} = \chi_c\,^c = 0$ to $\rho' = 0$, so that 
\[
\rho = \rho_* = const. > 0,
\]
equations  (\ref{coord-Omega-equ}) and (\ref{coord-T-trace-free-part})   imply 
$s = \frac{\epsilon}{2}\,\Omega - \frac{1}{8}\,\rho_*\,\Omega^2$,
$\Omega'' + \epsilon\,\Omega - \frac{1}{2}\,\rho_*\,\Omega^2 = 0$
and equations (\ref{coord-s-equ}), (\ref{coord-T-trace}), (\ref{coord-T-trace-free-part}) give
$s' = \frac{\epsilon}{2}\,\Omega' - \frac{1}{4}\,\rho_*\,\Omega\,\Omega'$,
which is satisfied by the function $s$ given above. The equations for $s$ are redundant under the given assumptions. So we are left with the initial value problems
\[
\Omega'' + \epsilon\,\Omega - \frac{1}{2}\,\rho_*\,\Omega^2 = 0,
\quad \Omega(0) = 0, \quad \Omega'(0)  = - \nu,
\]
which clearly have a smooth solutions near $\{\tau = 0\} = {\cal J}^+$. Where $\Omega' \neq 0$ (thus in particular near ${\cal J}^+$.) the ODE is equivalent to 
$(3\,\Omega'^2 + 3\,\epsilon\,\Omega^2 - \rho_*\,\Omega^3)' = 0$,
which implies with the boundary conditions 
\begin{equation}
\label{conf-first-order-FRW-equ}
3\,\Omega'^2 + 3\,\epsilon\,\Omega^2 - \rho_*\,\Omega^3 = \lambda.
\end{equation}
The decreasing solutions to this equation cover all the expanding ends of the FRW-type solutions.
With  (\ref{c-coord-conf-transf}) the usual (physical) equations (see \cite{hawking:ellis}) for $a(t)$ are implied by (\ref{conf-first-order-FRW-equ}).

\section{The subsidiary system}
\label{subs-syst}

To show that solutions to the reduced equations for data which satisfy the constraints do indeed satisfy the complete set of conformal field equations, it has to be shown that the  {\it zero quantities} $N_j$
and 
\begin{equation}
\label{zero-quantities} 
T_i\,^j\,_k, \quad \Delta^i\,_{jkl}, \quad A, \quad B_j, \quad C_{jl}, \quad D_j, \quad H_{jkl},  \quad F_{jkl},   
\end{equation}
vanish as a consequence of the reduced equations and the given initial data. Here 

\[
N_j \equiv \Omega\,f_j + U^k \Sigma_k\,U_j 
+  \Sigma_j + \Omega\,\,U^k f_k\, U_j,
\] 

\[
T_i\,^k\,_j\,e_k \equiv
- [e_i,e_j] + (\Gamma_i\,^l\,_j - \Gamma_j\,^l\,_i)\,e_l,
\]

\begin{equation}
\label{Delta-ijkl-def}
\Delta^i\,_{jkl} \equiv R^i\,_{jkl} - \Omega\,W^i\,_{jkl}
- 2\,\{g^i\,_{[k}\,L_{l] j} + L^i\,_{ [k}\,g_{l] j}\},
\end{equation}
with
\[
R^i\,_{jkl} = \Gamma_l\,^i\,_{j,\,\mu}\,e^{\mu}\,_k - \Gamma_k\,^i\,_{j,\,\mu}\,e^{\mu}\,_l
+ 2\,\Gamma_{[k}\,^{i\,p}\,\Gamma_{l]pj}
- 2\,\Gamma_{[k}\,^p\,_{l]}\,\Gamma_p\,^i\,_j,
\]

\[
A \equiv 6\,\Omega\,s - 3\,\Sigma_i\,\Sigma^i - 
\lambda - \frac{1}{4}\,\Omega^3\,\rho,
\]
\[
B_k \equiv \nabla_k\Omega - \Sigma_k
\]
\[
C_{jk} \equiv \nabla_{j}\,\Sigma_{k} + \,\Omega\,L_{j k} - s\,g_{j k}
- \frac{1}{2}\, 
 \Omega^2\,\rho\left(U_{j}\,U_{k} + \frac{1}{4}\,g_{j k}\right),
\]
\[
D_k \equiv \nabla_{k}\,s + \Sigma^{i}\,L_{ik} 
- \frac{1}{2}\,\Omega\,\rho\,\Sigma^{i}\left(U_{i}\,U_{k} + \frac{1}{4}\,g_{i k}\right)
- \frac{1}{8}\,\Omega\,\rho\,\Sigma_{k} - \frac{1}{24}\,\Omega^2\,\nabla_{k}\,\rho,
\]

\[
H_{jkl} \equiv \nabla_{k}\,L_{l j} - \nabla_{l}\,L_{k j} -  K_{jkl},
\]

\[
F_{jkl} \equiv \nabla_{i}\,W^{i}\,_{j k l}  - M_{jkl},
\]
where
\begin{equation}
\label{def-Mjkl}
M_{j k l} 
= \nabla_{[k} \rho\,\,U_{l]} \,U_{j} 
+ \frac{1}{3}\,\nabla_{[k} \rho \,\,g_{l] j}
\end{equation}
\[
+ \rho\,\,(\nabla_{[k}\,U_{l]} \,U_{j} 
+  U_{[l} \,\nabla_{k]}\,U_{j}
- f_{[k}\,g_{l]j} - 2\,f_{[k}\,U_{l]}\,U_j - U_{[k}\,g_{l]j}\,U^i\,f_i),
\]

\begin{equation}
\label{def-Fjkl}
K_{jkl} =  \Sigma_{i}\,W^{i}\,_{j k l} + \Omega\,M_{jkl}.
\end{equation}

\vspace{.1cm}

Some of these quantities vanish trivially because of symmetries, gauge conditions, or the reduced equations.  The latter  comprise equations (\ref{first-g-conf-geod-equ}),  (\ref{second-g-conf-geod-equ}), (\ref{rho-evol})
and 
\begin{equation}
\label{a-reduced-equations}
U^i\,T_i\,^k\,_j = 0, \quad 
U^k \Delta^i\,_{jkl}  = 0, \quad 
U^j B_j = 0, \quad U^j C_{jl} = 0, \quad U^j D_j = 0,  
\end{equation}

\begin{equation}
\label{b-reduced-equations}
H^j\,_{ja} = 0, \quad  H_{a0b} + H_{b0a} = 0,\quad a, b = 1, 2, 3, \quad  P_{ij} = 0, \quad 
Q_{ij} = 0,
\end{equation}
The zero quantities not in this list correspond to constraints or gauge conditions. Concerning the second of equations (\ref{a-reduced-equations}) we refer to the remarks below.

\vspace{.1cm}

In the following we shall use the covariant derivative operator $\nabla_j$ defined by the connection coefficients $\Gamma_i\,^j\,_k$ that satisfy  the gauge conditions and the reduced 
equations. This operator is metric in the sense that $\nabla_i\,g_{jk}  = 0$ but, as seen from
the first of conditions  (\ref{a-reduced-equations}), it is not known a priori whether the connection is  torsion free. 
In the following arguments will be needed the commutators of covariant  derivatives, which are for a function $\phi$ and a vector field $X^i$ in the case of a general metric connection of the form
\[
(\nabla_i\,\nabla_j  - \nabla_j\,\nabla_i)\,\phi = - T_i\,^l\,_j\,\nabla_l\,\phi
\]
\[
(\nabla_i\,\nabla_j  - \nabla_j\,\nabla_i)\,X^k = R^k\,_{lij}\,X^l - T_i\,^l\,_j\,\nabla_l\,X^j.
\]
To avoid carrying along various non-illuminating terms involving components of the torsion tensor we shall refer to such terms in an equation often in the form $\ldots + P(T)$, where the dots indicate the equation of interest and $P(T)$ is a generic symbol for a polynomial   in the components of the torsion tensor that satisfies $P(0) = 0$.
The equation above will then take the form
 \[
(\nabla_i\,\nabla_j  - \nabla_j\,\nabla_i)\,X^k = R^k\,_{lij}\,X^l + P(T).
\]
The other zero quantities in the list (\ref{zero-quantities}) will be kept  explicitly
 in an equation if needed to indicate how the calculations goes, otherwise the equations will be written  in the form
$\ldots + P(Z)$, where the dots indicate the members  of interest and $P(Z)$ is a polynomial in the components of the zero quantities (that may occasionally absorb a $P(T)$) with smooth coefficients that satisfies $P(0) = 0$.

\vspace{.1cm}

The regular system has been obtained from the original version of the conformal field equations by using 
the gauge requirements $N_j = 0$ and $\nabla_aN_b = 0$.  It needs to be shown that they are preserved by the reduced equations to establish that the original version of the conformal field equations is satisfied. They are needed in particular to show that  the equations for $\zeta_{ab}$ and $\xi$ imply the equations $U^i\Delta^0\,_{aib}  = 0$, $U^i\Delta^a\,_{0ib}  = 0$. The zero quantity $N_j$ plays a particular role because its vanishing follows directly from the reduced equations and the initial conditions.

\vspace{.1cm}

\noindent
{\it  If $N_k = 0$ on a hypersurface transverse to the flow lines of $U^{k}$ (which will, for instance, be the case if data are prescribed on $\{ \Omega = 0\}$), this relation
is preserved along the flow lines of $U$ as a consequence of the reduced equations.
}

\vspace{.1cm}

\noindent
In fact, equations (\ref{first-g-conf-geod-equ}) and (\ref{second-g-conf-geod-equ})
imply
\[
\nabla_U N_{i} = U^k\,U^l\,C_{kl}\,U_i + U^k\,C_{ki} + U^kB_k\,(f_i \,+ U^l f_l\,U_i)
- U_{i} \,f^{k} N_{k},
\]
which reduces with (\ref{a-reduced-equations}) to the linear homogeneous ODE 
\begin{equation}
\label{N-subs-equ}
\nabla_U N_{i} = - U_{i} \,f^{k} N_{k},
\end{equation}
along the flow lines of $U$. From this the assertion follows. Since the solution to the reduced equations is ruled by the flow lines it follows also that $\nabla_iN_j = 0$ on the solution.

\vspace{.1cm}

\noindent
It can thus be assumed that $N_j = 0$, $\nabla_aN_b = 0$ so that we have indeed 
$U^i\Delta^0\,_{aib}  = 0$ and the equivalent equation $U^i\Delta^a\,_{0ib}  = 0$ as written in  (\ref{a-reduced-equations}).

\vspace{.1cm}

The subsequent discussion follows to some extent the derivation of subsidiary systems 
in earlier work on the conformal field equations. 
It will be convenient to use for the covariant derivative of a given tensor field $X_{ij}\,^k$  the notation
\[
\nabla_l X_{ij}\,^k =  e_l(X_{ij}\,^k) + (\Gamma X)_{lij}\,^k,
\]
so that $X_{ij}\,^k \rightarrow (\Gamma X)_{lij}\,^k$ denotes a purely algebraic  linear operator 
which does not involve derivatives.

\vspace{.1cm}

The connection defined by the $\Gamma_i\,^j\,_k$ and the associated
torsion and curvature
tensor satisfy the first Bianchi identity
\[
\sum_{(jkl)} \nabla_j\,T_k\,^i\,_l =
\sum_{(jkl)} (R^i\,_{jkl} + T_j\,^m\,_k\,T_l\,^i\,_m),
\]
where $\sum_{(jkl)}$ denotes the sum over the cyclic permutation of the
indices $jkl$.
Setting here $j = 0$, observing that the symmetries of $C^i\,_{jkl} = \Omega\,W^i\,_{jkl}$ and
$L_{kl}$ imply
$\sum_{(jkl)} R^i\,_{jkl} = \sum_{(jkl)} \Delta^i\,_{jkl}$ and taking
into account the
reduced equations, we get from this the equation
\begin{equation}
\label{T-subs-equ}
\nabla_0\,T_k\,^i\,_l  = - (\Gamma T)_{l0 }\,^i\,_k  + (\Gamma T)_{k0}\,^i\,_l 
+ 3\,\sum_{(0kl)} (\Delta^i\,_{0kl} + T_0\,^m\,_k\,T_l\,^i\,_m) = P(Z).
\end{equation}

\vspace{.1cm}

To obtain an equation of the desired type for  $\Delta^i\,_{jkl}$ we show that the right hand side of the 
identity
\[
\nabla_j \Delta^i\,_{mkl} + \nabla_l \Delta^i\,_{mjk} + \nabla_k \Delta^i\,_{mlj} = \frac{1}{2}\,
\epsilon_{njkl}\,\epsilon^{npqr}\,\nabla_p \Delta^i\,_{mqr}.
\]
can be written as a linear expression in the zero quantities. We write (\ref{Delta-ijkl-def}) in the form
\[
R^i\,_{jkl} =
\Delta^i\,_{jkl} +  \Omega\,W^i\,_{jkl} + G^i\,_{jkl} + E^i\,_{jkl},
\]
with
\[
G^i\,_{jkl} = L\,g^i\,_{[k}\,g_{l]j}, \quad  L = L_i\,^i, \quad 
E^i\,_{jkl} = 2\,\{g^i\,_{[k}\,L^*_{l] j} + L^{*i}\,_{ [k}\,g_{l] j}\}, 
\quad L^*_{l j} = L_{l j} - \frac{1}{4}\,L\,g_{l j},
\]
and  use the second Bianchi identity
\begin{equation}
\label{2Bianchi}
\sum_{(jkl)} \nabla_j\,R^i\,_{mkl} =
- \sum_{(jkl)} R^i\,_{mpj}\,T_k\,^p\,_l,
\end{equation}
to obtain
\[
\epsilon^{njkl}\,\nabla_j \Delta_{imkl} = - \epsilon^{njkl}\,\left(\nabla_j\Omega\,W_{imkl} 
+ \Omega\,\nabla_j W_{imkl}  
\right.
\]
\[
\left.
+ \nabla_jG_{imkl} + \nabla_jE_{imkl} + R_{impj}\,T_k\,^p\,_l\right).
\]
The well known facts that the left and right duals of $W_{ijkl}$ and $G_{ijkl}$ coincide respectively 
while the left dual of $E_{ijkl}$ differs from its right dual by a sign then imply with the reduced equations
\[
\epsilon_n\,^{jkl}\,\nabla_j \Delta_{imkl} = \epsilon_{im}\,^{kl}\,\left( \nabla_j\Omega\,W^{j}\,_{nkl} 
+ \Omega\,\nabla_j W^{j}\,_{nkl}  
\right.
\]
\[
\left.
+ \nabla_jG^{j}\,_{nkl} - \nabla_jE^{j}\,_{nkl} \right)  - \epsilon_n\,^{jkl}\,R_{impj}\,T_k\,^p\,_l
\]
\[
= \epsilon_{im}\,^{kl}\,\left( \nabla_j\Omega\,W^{j}\,_{nkl} 
+ \Omega\,(F_{nkl} + M_{nkl})  
\right.
\]
\[
\left.
2\,\nabla_{[k}L\,g_{l]n} - 2\,\nabla_{[k}\,L_{l]n} - 2\,\nabla_j\,L^j\,_{[k}\,g_{l]n} \right)  - \epsilon_n\,^{jkl}\,R_{impj}\,T_k\,^p\,_l
\]
\[
= \epsilon_{im}\,^{kl}\,\left( \nabla_j\Omega\,W^{j}\,_{nkl} 
+ \Omega\,(F_{nkl} + M_{nkl})  
\right.
\]
\[
\left.
- H_{nkl} 
- \Sigma_{i}\,W^{i}\,_{n k l} - \Omega\,M_{nkl}
- 2\,H^j\,_{j[k}\,g_{l]n}\right)  - \epsilon_n\,^{jkl}\,R_{impj}\,T_k\,^p\,_l. 
\]
In the last step it has been used that $K^j\,_{jl} = 0$. This follows because the tensor $W^i\,_{jkl}$ has vanishing contractions and because equations
(\ref{first-g-conf-geod-equ}) and ({\ref{rho-evol}), which are satisfied as members of the reduced system, imply that $M^j\,_{jl} = 0$.
Using again the reduced equations we finally get

\begin{equation}
\label{Delta-imkl-subs}
\nabla_{0} \Delta^i\,_{mkl} = 
- (\Gamma \Delta)_l\,^i\,_{m0k} 
+ (\Gamma \Delta)_k\,^i\,_{m0l} 
\end{equation}
\[
- \frac{1}{2}\,\epsilon^n\,_{0kl}\left\{
\epsilon^i\,_{m}\,^{kl}\,\left(B_p\,W^{p}\,_{nkl} 
+ \Omega\,F_{nkl}  
- H_{nkl} 
- 2\,H^p\,_{pk}\,g_{l n}\right)  - \epsilon_n\,^{jkl}\,R^i\,_{mp0}\,T_k\,^p\,_l
\right\} = P(Z).
\]

\vspace{.1cm}

A direct calculation gives for the quantity
\begin{equation}
\label{alg-zero-q}
A = 6\,\Omega\,s - 3\,\Sigma_i\,\Sigma^i - 
\lambda - \frac{1}{4}\,\Omega^3\,\rho,
\end{equation}
the relation
\[
\nabla_j A = 6\,\Omega\,D_j - 6\,\Sigma^i\,C_{ji} + (6\,s - \frac{3}{4}\,\Omega^2\,\rho)\,B_j.
\]
On the initial slice, where the zero quantities on the right hand side vanish by the construction  of the initial data, this relation reduces to $\nabla_j A = 0$. 
This implies that $A = 0$ on that slice if it holds at one point of it. In the case of `physical' data (i. e. 
$\Omega = 1$) the condition $A = 0$ reduces to $0 = 4\,\hat{A} = \hat{R} - 4\,\lambda - \hat{\rho}$, which will be satisfied by the construction of the physical data. Using the freedom to prescribe $\Omega$ and its time derivative on the initial slice the condition $A = 0$ can also be achieved in the transition to conformal data. We recall that  the relation $A = 0$ served to  determine the value of $\Sigma_j$
in our discussion of the conformal data on $\{\Omega = 0\}$.  
With the reduced equations the relation above implies that
\[
\nabla_U A = 0.
\]
{\it We can thus assume that $A = 0$ on the solution manifold}.

\vspace{.1cm}

A straightforward but lengthy calculation shows that
the fields
\[
Z^B_{jk} = \nabla_{[j}\,B_{k]}, \quad Z^C_{jkl} = \nabla_{[j}\,C_{k]l}, \quad Z^D_{jk} = \nabla_{[j}\,D_{k]}, 
\]
can be expressed as linear (homogeneous) functions of the zero quantities with smooth coefficients. Taking into account  the reduced equation $U^j B_j = 0$, 
$U^j C_{jl} = 0$, $U^j D_j = 0$ one gets
\[
U^j\,\nabla_{j}B_{k} = 2\,U^j\,Z^B_{jk} + U^j\,\nabla_{k}\,B_{j} 
= 2\,U^j\,Z^B_{jk} +  \nabla_{k}\,(U^jB_{j}) - (\nabla_{k}U^j)\,B_{j} = P(Z). 
\]
Similar calculations give
\begin{equation}
\label{B-C-D-subs}
U^j\,\nabla_{j}B_{k} = P(Z),  \quad \quad 
U^j\,\nabla_{j}C_{kl} = P(Z), \quad \quad 
U^j\,\nabla_{j}D_{k} = P(Z).
\end{equation}

The remaining subsidiary equations are obtained by analyzing the expressions 
\[
\nabla_{[l}\,H^i\,_{jk]}
\quad \quad \nabla^j\,F_{jkl}, 
\]
from two different points of view. As a preparation we observe the algebraic relations 
\begin{equation}
\label{M-algebra}
M_{jkl} = - M_{jkl}, \quad \quad  M_{[jkl]} = 0, \quad \quad M^j\,_{jl} = 0.
\end{equation}
The first of them follow immediately from the definition while, as pointed out above, the last one follows as a consequence of the reduced equations (\ref{first-g-conf-geod-equ}) and ({\ref{rho-evol}). These relations imply
\begin{equation}
\label{F-algebra}
F_{jkl} = - F_{jkl}, \quad \quad  F_{[jkl]} = 0, \quad \quad F^j\,_{jl} = 0,
\end{equation}
and also
\begin{equation}
\label{K-algebra}
K^j\,_{jl} = 0.
\end{equation}
Moreover, a straightforward though fairly lengthy calculation which makes  repeatedly use of the reduced equations, shows that 
\begin{equation}
\label{nabla-M-rel}
\nabla^jM_{jkl} = P(Z),
\end{equation}
and 
\[
\nabla_l K^l\,_{jk} = \nabla_l \Sigma_i\,W^{il}\,_{jk} + \Sigma_i\,\nabla_lW^{il}\,_{jk}
+ \nabla_l\Omega\,M^l\,_{jk} + \Omega\,\nabla_l M^l\,_{jk}
\]
\[
= C_{li}\,W^{il}\,_{jk} + B_l M^l\,_{jk} - \Sigma_l F^l\,_{jk} + \Omega\,\nabla_l M^l\,_{jk} = P(Z).
\]
From this follows the relation 
\begin{equation}
\label{nabla-H-contraction}
\nabla_{[l} H^l\,_{jk]} =\Delta^l\,_{p[lj}\,L_{k]}\,^p - \nabla_{[l} K^l\,_{jk]} + P(T) = P(Z).
\end{equation}
Similar calculations, which use that the left and right duals of the conformal Weyl tensor coincide,  gives
\[
\epsilon^{qljk}\,\nabla_l H^p\,_{jk} = \epsilon^{qljk}\,(
\nabla_l\nabla_jL_k\,^p - \nabla_l(\Sigma_n\,W^{np}\,_{jk} + \Omega\,M^p\,_{jk}))
\]
\[
= \epsilon^{qljk}\,(\Delta^p\,_{nlj}\,L_k\,^n + W^p\,_{nlj}\,C_k\,^n - B_l\,M^p\,_{jk}) + 
\epsilon^{npik}\,\Sigma_n\,F^q\,_{jk}
\]
\[
+ \frac{1}{2}\,\rho\,\Omega^2\,\epsilon^{qljk}\,\,W^p\,_{njk}\,U^n\,U_l
- 2\,\Sigma_l\,M^{(q}\,_{jk}\,\epsilon^{p)ljk} - \Omega\,\nabla_l\,M^p\,_{jk}\,\epsilon^{pljk} .
\]
From equations  (\ref{M-algebra}), (\ref{nabla-M-rel}) follows that
\[
\epsilon_{pqmn}\, \nabla_l\,M^p\,_{jk}\,\epsilon^{qljk}  = P(Z).
\]
Solving  the equation $N_l = 0$ for $\Sigma_l$ and inserting this into the equation above, we thus finally get

\[
\epsilon^{qljk}\,\nabla_l H^p\,_{jk} = 
 \frac{1}{2}\,\rho\,\Omega^2\,\epsilon^{qljk}\,\,W^p\,_{njk}\,U^n\,U_l
 \]
 \[
- 2\,\Sigma_l\,M^{(q}\,_{jk}\,\epsilon^{p)ljk} - \Omega\,\nabla_l\,M^{(p}\,_{jk}\,\epsilon^{q)ljk} + P(Z).
\]
\[
\epsilon^{qljk}\,\nabla_l H^p\,_{jk} = 
 \frac{1}{2}\,\rho\,\Omega^2\,\epsilon^{qljk}\,\,W^p\,_{njk}\,U^n\,U_l
+ 2\,\,\nabla_U\Omega\,\,U_l\,M^{(q}\,_{jk}\,\epsilon^{p)ljk} 
\]
 \[
+  \Omega\,\{
 2\,(f_l \,+ <U,f>U_l)\,M^{(q}\,_{jk}\,\epsilon^{p)ljk} 
- \nabla_l\,M^{(p}\,_{jk}\,\epsilon^{q)ljk}\} + P(Z).
\]
A direct calculation shows now that 
\begin{equation}
\label{H00-Hab-rel}
\epsilon^{0ljk}\,\nabla_l H^0\,_{jk} = P(Z), \quad \quad 
\epsilon^{aljk}\,\nabla_l H^b\,_{jk} = P(Z), \quad a, b = 1, 2, 3.
\end{equation}

\vspace{.1cm}

After solving the $9$ reduced equations for the components $L_{0a}$, $L_{ab}$, they  
resume their original form if $1/6\,R$ is replaced again by $L_j\,^j$.
To show that they imply for suitably given initial data the full set $H_{jkl} = 0$, it needs to be shown that 
\[
H_{abc} = 0, \quad \quad H_{0ab} = 0, \quad a  \neq b.
\]
In fact, the equation $0 = H^j\,_{ja} = - H_{00a} + g^{cd}\,H_{cad}$ implies then that $H_{00a} = 0$ and with the identities 
\[
H_{jkl}  = - H_{jlk}  \,\,\, \mbox{and} \,\,\,    \epsilon^{ijkl}\,H_{jkl} = 0, \,\,\, \mbox{i.e.}\,\,\, 
\epsilon^{abc}\,H_{abc} = 0 \,\,\, \mbox{and} \,\,\, H_{0ab} + H_{b0a} + H_{ab0} = 0, \,\,\, a \neq b,
\]
and the  reduced equation $H_{a0b} + H_{b0a} = 0$ it follows then that
\[
0 = H_{0ab} = -  H_{b0a} + H_{a0b} =  2\, H_{a0b} \quad a \neq b,
\]
which exhaust the remaining cases. 

\vspace{.1cm}

We derive now the equations for the zero quantities above.
The reduced equation $H^j\,_{ja} = 0$ implies that 
$\nabla_k H^l\,_{la} = (\Gamma\,H)_k\,^i\,_{la} = P(Z)$. Observing this in equations 
(\ref{nabla-H-contraction}) we an equation of the form
\begin{equation}
\label{1-H-red-equ}
\nabla_0 H_{0ab} - g_{cd}\,\nabla_cH_{dab} = P(Z).
\end{equation}
On the other hand we have by (\ref{H00-Hab-rel}) 
\[
\nabla_0 H_{dab} + \nabla_b H_{d0a} - \nabla_aH_{d0b} = 3\,\nabla_{[0} H_{|d| ab]} = P(Z)
\]
and 
\[
\nabla_d H_{0ab} + \nabla_b H_{0da} + \nabla_aH_{0bd} = 3\,\nabla_{[d} H_{|0|ab]} = P(Z)
\]
(where  indices with a modulus sign are exempt from the anti-symmetrization).
Observing  the relations $H_{a0b} = - H_{b0a}$ and $2\,H_{c0d} = H_{0cd}$ implied be the reduced equations, one gets from this an equation of the form
\begin{equation}
\label{2-H-red-equ}
2\,\nabla_0 H_{dab} - \nabla_d H_{0ab} = P(Z).
\end{equation}
Equations (\ref{1-H-red-equ}), (\ref{2-H-red-equ}) constitute  a system of equations for the unknowns $H_{0ab}$ and $H_{abc}$ which is, for given right hand sides, symmetric hyperbolic.

\vspace{.1cm}

The properties  (\ref{F-algebra}) imply in particular the relation $F^a\,_{0a} = F^i\,_{0i} = 0$. The 
field $P_{ij}$ and $Q_{kl}$ introduced in 
(\ref{P-ij-def}} and (\ref{Q-ij-def}) are thus completely represented by 
\[
P_{ab}  =  - F_{(a |0| b)}, \quad \quad Q_{ab} =  - \frac{1}{2}\,F_{(a}\,^{cd}\,\epsilon_{b)cd}.
\]
To discuss the remaining content of the field $F_{jkl}$ we recall the definitions
\[
P_a = F_{0a0}, \quad \quad  
Q_b = - \frac{1}{2}\,F_{0cd}\,\epsilon_b\,^{cd}, 
\]
given in the discussion of the constraints. These fields exhaust the information in $F_{0a0}$ and $F_{0bc}$. Because $F_{a0b}$ is trace free it remains to control its anti-symmetric part. The relation  $F_{[jkl]} = 0$  gives 
\[
- Q_c\,\epsilon^c\,_{ab} = \frac{1}{2}\,F_{0de}\,\epsilon_c\,^{de}\,\epsilon^c\,_{ab} 
= F_{0ab} = F_{a0b} - F_{b 0 a},
\]
whence 
\[
F_{a0b} = - P_{ab} - \frac{1}{2}\,\epsilon_{abc}\,Q^c.
\]
Because $F_{abc}\,\epsilon^{abc} = 0$, the field $F_{acd}\,\epsilon_{b}\,^{cd}$ is trace free. Contracting its anti-symmetric part suitably twice with epsilons and using that gives $F^j\,_{jl} = 0$ gives
\[
F_{[a}\,^{cd}\,\epsilon_{b]cd}  =  - F_d\,^{dc}\,\epsilon_{cab} = - F_{00c}\,\epsilon^c\,_{ab}
= P_c\,\epsilon^c\,_{ab}, 
\]
and thus 
\[
F_{abc} = \frac{1}{2}\,Q_{ad}\,\epsilon_{bc}\,^d - h_{a[b}\,P_{c]}.
\]
Observing now the reduced equations $P_{ab}  =  0$ and $Q_{ab} = 0$, the remaining content of 
$F_{jkl}$  is then described by the formula 
\[
F_{jkl} = 3\,U_j\,P_{[k}\,U_{l]} - g_{j[k}\,P_{l]} 
+ Q_i\,(U_j\,\epsilon^i\,_{kl} - \epsilon^i\,_{j[k}\,U_{l]}).
\]
Inserting this into $\nabla^j\,F_{jkl}$ and projecting suitably gives his  
\[
 (\nabla_U\,P_l)\,h^l\,_i + \frac{1}{2}\,\epsilon_i\,^{kj}\,\nabla_k\,Q_j = \nabla^j\,F_{jkl}\,U^k\,h^l\,_i 
+ P(Z),
\]
\[
 (\nabla_U\,Q_l)\,h^l\,_i 
- \frac{1}{2}\,\epsilon_i\,^{kj}\,\nabla_k\,P_j = \frac{1}{2}\,\nabla^j\,F_{jkl}\,\epsilon_i\,^{kl} + P(Z).
\]
Working then out $\nabla^j\,F_{jkl}$ explicitly and observing (\ref{nabla-M-rel})
one finally gets equations of the form
\begin{equation}
\label{P-subs-equ}
P_{a, \,0} + \frac{1}{2}\,\epsilon_a\,^{bc}\,D_b\,Q_c = P(Z),
\end{equation}
\begin{equation}
\label{Q-subs-equ}
Q_{a,\,0}  - \frac{1}{2}\,\epsilon_a\,^{bc}\,D_b\,P_c = P(Z).
\end{equation}
For given right hand sides this is a symmetric hyperbolic system for the fields $P_a$ and $Q_a$.

\vspace{.2cm}

We have seen above that solutions to the reduced equations for  suitably arranged initial data
satisfy $N_j = 0$ and $A = 0$.
Equations (\ref{T-subs-equ}), (\ref{Delta-imkl-subs}), (\ref{B-C-D-subs}),
(\ref{1-H-red-equ}), (\ref{2-H-red-equ}), (\ref{P-subs-equ}), (\ref{Q-subs-equ}) constitute a system of differential equations for those of the remaining components of the zero quantities (\ref{zero-quantities}) 
which do not vanish already because of gauge conditions or the reduced equation. The system is symmetric hyperbolic and has characteristics which are time-like or null with respect to the metric
$g_{\mu\nu}$ that is supplied by the reduced system.

\vspace{.2cm}

\noindent
{\it It follows that a solution to the reduced system for data that satisfy the conformal constraints on the initial slice satisfies on the domain of dependence of the initial slice  the gauge conditions and the complete set of conformal Einstein-$\lambda$-dust equations.}

\section{Existence and strong future stability}
\label{ex-strong-stab}

In this section the properties of the conformal field equations derived above  and standard results about quasi-linear symmetric hyperbolic systems will be used to draw conclusions on the global structure of solutions to the  Einstein-$\lambda$-dust equations.
Since we are mainly  interested in $C^{\infty}$ solutions and not in the weakest possible smoothness assumptions on the data we refrain from specifying Sobolev norms.
We refer to \cite{friedrich:1991} for details of the patching arguments in the context of Cauchy stability and for some relevant PDE reference.

\subsection{Existence of asymptotically simple solutions}

To construct solutions to the Einstein-dust equations with positive cosmological constant 
$\lambda$ that admit a smooth conformal boundary in their infinite future we consider Cauchy problems for the reduced field equations on $\mathbb{R} \times S$ where data are prescribed on the submanifold $\{0\} \times S$. We identify the latter diffeomorphically
with the manifold $S$ underlying a given {\it asymptotic end data set} as considered in section 
\ref{as-end-dat}. The {\it conformal time} variable $\tau$ in the reduced field equations will correspond to the factor $\mathbb{R}$ above and it will be assumed that $\tau = 0$ on $S$. 
The conformal gauge source function represented by the Ricci scalar $R[g]$ of the conformal metric $g$ to be constructed will be required to vanish and it is assumed that the condition $R[g] = 0$ is also underlying the construction of the given asymptotic end data. 
A fixed gauge source function will in general only work well for some limited time. For our purpose this will suffice, however, because it will be arranged that a finite  interval of the conformal time 
$\tau$ will cover an interval of  {\it physical} time of infinite extent.

\vspace{.1cm}

Since $S$ is compact and may have complicated topology, we use the fact that the hyperbolicity of the reduced equation allows us  to obtain a solution on a neighborhood of
$S \sim \{0\} \times S$ in $\mathbb{R} \times S$ by patching together local solutions.
Compactness implies that $S$ can be covered by a finite number of open subsets 
$V_A$, $A = 1, 2, \ldots , k$, of $S$  which carry 
smooth local coordinates $x ^{\alpha}$, $\alpha = 1, 2, 3$,  and a smooth frame field $e_a$, $a = 1, 2, 3$, that satisfies $h_{ab} \equiv h(e_a, e_b) = \delta_{ab}$, where $h$ denotes the 3-metric on $S$ supplied by the asymptotic end data. It can be assumed that their exist shrinkings $V'_A$ 
with compact closure $\overline{V'}_A$ in $V_A$ so that the $V'_A$ still define an open covering 
and the boundary of $V'_A$ in $V_A$ is smooth. Standard results on symmetric hyperbolic systems then imply the existence of smooth solutions  to the reduced field equations on open neighbourhoods 
${\cal D}_A$ of $V'_A$ in $\mathbb{R} \times S$ which imply  on $V'_A$ the data induced on 
 $V'_A$ by the asymptotic end data on $S$  in the gauge chosen on $V_A$. It can be assumed that the solution  
 extends smoothly to the closure of ${\cal D}_A$ in $\mathbb{R} \times S$ with 
 $\det(e^{\mu}\,_k) \neq 0$ so that  
${\cal D}_A$ acquires a boundary which consists of (i) smooth hypersurfaces ${\cal H}^{\pm}_A$
in the future/past of ${\cal D}_A$ which are null with respect to the solution metric $g$ 
and approach
$\overline{V'}_A \setminus V'_A$ in their past/future, (ii) the intersection of $\overline {{\cal D}_A}$ 
with hypersurfaces $\{\tau = \tau_{\pm}\}$ in $\mathbb{R} \times S$
defined by some constants $\tau_- < 0 < \tau_+$ (which can be chosen to be the same for all
$V'_A$), and (iii) the three 2-dimensional 
edges diffeomorphic to $\overline{V'}_A \setminus V'_A$ where these hypersurfaces approach each other. It can be assumed that the solution on ${\cal D}_A$ is globally hyperbolic 
with respect to metric $g$.
The subsidiary system then implies that the full set of conformal 
Einstein-$\lambda$-dust equations is satisfied on $\overline {{\cal D}_A}$.

If $p \in V'_A \cap V'_B$, there exists an open neighborhood $V_p \subset V'_A \cap V'_B$ of $p$
so that solutions are given in the domain of dependence ${\cal D}_{A, p}$ of $V_p$ in ${\cal D}_A$ 
as well as in the domain of dependence ${\cal D}_{B, p}$ of $V_p$ in ${\cal D}_B$. On $V_p$ these two  solutions can be related to each other because the coordinate and frame transformations which relate the data induced on $V_p$ by the data on $V'_A$ 
and the data on $V'_B$ respectively  are known explicitly. Because the gauge inherent in  the reduced equations is evolved by invariant propagation laws along the invariantly defined flow lines of the flow field $U$, the coordinate and frame transformations 
extend, independent of $\tau$, and allow us to relate the solution on ${\cal D}_{A, p}$ isometrically to the solution on 
${\cal D}_{B, p}$. By extending the argument it follows that the solution induced on the domain of dependence of 
$V'_A \cap V'_B$ in ${\cal D}_A$ can be identified isometrically with the solution induced on the domain of dependence of $V'_A \cap V'_B$ in ${\cal D}_B$. 

By patching together the local solutions,
we obtain a smooth, globally hyperbolic solution to the 
conformal Einstein-$\lambda$-dust equations
on a subset of the form
$M = [\tau_*, \tau_{**}] \times S$ of $\mathbb{R} \times S$ 
with constants $\tau_* < 0 < \tau_{**}$ so that the conformal factor obtained on $M$ satisfies  
$\Omega > 0$ on 
$\hat{M} = [\tau_*, 0[ \times S$
while $\Omega < 0$ on $\check{M} = ]0, \tau_{**}] \times S$.
\\

The 
 hypersurfaces $S_{\sigma} = \{\tau = \sigma = const.\}$ with $\tau_* \le \sigma \le \tau_{**}$ can be required to be space-like. In fact, with  the co-normal to $\{\tau = const.\}$ given by 
$n_{\mu} = - a\,\tau_{,\mu}$
the future directed  normal is given by
\[
n^{\mu} = \frac{ - a\,g^{\mu 0}}{\sqrt{|a^2\,g^{00}|}} = - \frac{\eta^{jk}\,e^{\mu}\,_j\,e^0\,_k}
{\sqrt{|\eta^{jk}\,e^{0}\,_j\,e^0\,_k|}} 
= \frac{\delta^{\mu}\,_0 - \eta^{ab}\,e^{\mu}\,a\,e^0\,_b}{\sqrt{1 - \eta^{ab}\,e^{0}\,_a\,e^0\,_b}},
\]
and the condition $n_{\mu}\,n ^{\mu} = - 1$ implies the expression
\begin{equation}
\label{2nd-a-form}
a = \frac{1}{\sqrt{1 - \eta^{ab}\,e^0\,_a\,e^0\,_b}}.
\end{equation}
Moreover,
\begin{equation}
\label{2nd-n-form}
n^{\mu} 
= a\,(\delta^{\mu}\,_0 - \eta^{ab}\,e^{\mu}\,_a\,e^0\,_b)
= 
\frac{1}{a}\,(U^{\mu}
- u^{\alpha}\,\delta^{\mu}\,_{\alpha})
\quad \mbox{with} \quad u^{\alpha} = a^2\,\eta^{ab}\,e^{\alpha}\,_a\,e^0\,_b.
\end{equation}
We thus require that  
 \begin{equation}
 \label{2nd-e-restr}
 e^{0}\,_a\,e^{0}\,_b\,\eta^{ab} \le const. < 1  \quad \mbox{on} \quad M,
 \end{equation}
which can be achieved with suitable choices of $\tau_*$ and $\tau_{**}$ because $e^0\,_a = 0$ on $S_0$.  The 
 hypersurfaces $S_{\sigma} $ will then be Cauchy hypersurfaces for $(M, g_{\mu\nu})$. 
To simplify things, 
so that we only need to consider  the regularized reduced equations involving the unknowns 
$\zeta_{ab}$ and $\xi$,
it will also be assumed that $\Omega_{,\tau} < 0$ on $M$, which makes sense because $\Omega_{,\tau} = - \nu$ 
on $S_0$.

\vspace{.1cm}

The metric $g_{\mu\nu}$, the conformal factor $\Omega$, the flow field $U$ and the density function $\rho$ are then such that 
the `physical' fields
\begin{equation}
\label{conf-phys-field-rel}
\hat{g}_{\mu\nu} = \Omega^{-2}\,g_{\mu\nu}, \quad
\hat{U}_{\mu} = \Omega^{-1}\,U_{\mu}, \quad
\hat{\rho} = \Omega^{3}\,\rho
\end{equation}
define a solution to the Einstein-$\lambda$-dust equations on the manifold $\hat{M}$ with 
$\hat{\rho} \ge 0$ on $\hat{M}$. Extending smoothly to $S_{\tau_*}$, this solution 
admits an extension into the past of $S_{\tau_*}$ but we are not interested here in controlling something like a maximal globally hyperbolic solution. What is important for us is that 
the set ${\cal J}^+ \equiv S_0 = \{\Omega = 0\}$ defines for the 
solution $(\hat{M}, \hat{g}_{\mu\nu})$ a smooth conformal boundary at future time-like infinity.

\vspace{.1cm}

Equations (\ref{O-torsion-free condition}) to (\ref{f-rho-equ}) are invariant under the transformation
which implies the map
\[
\Omega \rightarrow - \Omega, \quad 
\nabla_k\Omega \rightarrow - \nabla_k\Omega, \quad s \rightarrow - s, \quad  W^i\,_{jkl} \rightarrow -  W^i\,_{jkl},  
\quad \rho \rightarrow - \rho, \quad \nabla_k\rho \rightarrow - \nabla_k\rho,
\]
but  leaves the fields $e^{\mu}\,_k$, $\Gamma_i\,^j\,_k$, $L_{jk}$, and $U^k$ unchanged.
It follows  that after performing this transition on $M$ and restricting to  $\check{M}$ 
gives us another solution to the Einstein-$\lambda$-dust equations on the manifold $\check{M}$. It follows, however, that then $\hat{\rho} \le 0$ on $\check{M}$. 
For this solution the set $ \{\Omega = 0\}$ defines a smooth conformal boundary in the infinite past. In this article we shall not be interested in this solution any further.

\vspace{.1cm}

Two facts have been used above to obtain solutions whose conformal structures extend smoothly across future time-like infinity so as to define there smooth conformal boundaries: (i) 
The Einstein-$\lambda$-dust equations admit conformal representations which imply with suitable gauge conditions systems of evolution equations that are hyperbolic irrespective of the sign of the conformal factor $\Omega$, (ii) 
some requirements needed to ensure the existence of smooth conformal extensions
{\it are put in by hand} by starting from asymptotic end data.

The case of the Nariai solution, an explicit, geodesically complete  solution to the Einstein-$\lambda$-dust equations 
with $\hat{\rho} = 0$, shows that that  the property (i) is by itself not sufficient to ensure the existence of a smooth conformal  boundary (see \cite{friedrich:beyond:2015}). This raises the question whether the use of asymptotic end data may result in the construction of a very restricted class of solutions.

The following argument, introduced in the vacuum case in \cite{friedrich:1986b} and used in the presence of conformally invariant matter fields in \cite{friedrich:1991},
shows that  the existence of smooth asymptotic conformal structures is in fact a fairly general feature of solutions to the Einstein-$\lambda$-dust equations. The smooth extensibility
of the conformal structure across future time-like infinity will
be {\it derived} as a consequence of the property (i) of the Einstein-$\lambda$-dust equations and the existence of a given reference solution that admits a smooth asymptotic structure.

\subsection{Strong future stability of the solutions}

Let
\begin{equation}
\label{B-unknnowns}
\Delta  = (e^{\mu}\,_k, \,\, \Gamma_i\,^j\,_k, \,\, \zeta_{ab}, \,\,\xi, \,\,
f_k, \,\, \Omega, \,\,\nabla_i\Omega, 
\,\, s, \,\, L_{jk}, \,\, W^i\,_{jkl}, \,\,
U^k, \,\, \rho), 
\end{equation}
be one  of the solutions constructed above. 
The associated physical fields $\hat{g}_{\mu\nu} = \Omega^{-2}\,g_{\mu\nu}$,
$\hat{U}^{\mu} = \Omega\,U^{\mu}$, $\hat{\rho} = \Omega^3\,\rho$ 
then induce on the Cauchy hypersurface $S' \equiv S_{\tau_*}$ with local coordinates 
$x^{\alpha}$, $\alpha = 1, 2, 3$,  
standard Cauchy data 
$\hat{\delta} = (\hat{h}_{\alpha \beta}, \,\hat{\kappa}_{\alpha \beta}, \,\hat{u}^{\alpha}, \,\hat{\rho})$,
i.e. a solution to the constraints (\ref{hat-Ham-constr}) and (\ref{hat-mom-constr}),
where $\hat{u}^{\alpha}$ denotes the orthogonal projection of $\hat{U}^{\mu}$ onto $S'$.

As a first step towards showing that the asymptotic simplicity of the solution above is preserved under sufficiently small perturbations  of the data $\hat{\delta}$, any given standard Cauchy data set  on $S'$ needs to be transformed into a suitable Cauchy data set  for the conformal field equations. This involves several transformations and a suitable handling of the gauge freedom which will be discussed now by showing how the restriction of $\Delta$ to $S'$ is obtained from $\hat{\delta}$.

Conformal data $\delta = (h_{\alpha \beta}, \,\kappa_{\alpha \beta}, \,u^{\alpha}, \,\rho)$  on $S'$ 
 are obtained from the standard data $\hat{\delta}$
by using the functions $\Omega > 0$ and $\nabla_U\Omega < 0$ on $S'$ to define
\[
h_{\alpha \beta} = \Omega^2\,\hat{h}_{\alpha \beta}, \quad 
u^{\alpha} = \Omega^{-1}\,\hat{u}^{\alpha}, \quad 
\rho = \Omega^{-3}\,\hat{\rho},
\]
and, using the transformation law of second fundamental forms under conformal rescalings,
\[
\kappa_{\alpha \beta} = \Omega\,(\hat{\kappa}_{\alpha \beta} 
+ \hat{h}_{\alpha \beta}\,\nabla_n\Omega).
\]
Here $n$ denotes the future directed unit normal to $S'$ with respect to $g$, which is related to the flow vector field $U$ and its projection $u$ onto $S'$ (that represents the shift vector field on $S'$, see the ADM representation of $g$ below) by the relation 
\[
n = \frac{1}{a}\,(U - u) \quad \mbox{with} \quad a = \sqrt{1 + h_{\alpha \beta}\,u^{\alpha}\,u^{\beta}},
\]
where the expression for the positive lapse function $a$ is obtained from
\[
- 1 = g(U, U) = a^2\,g(n, n) + g(u, u)
= - a^2 + h_{\alpha \beta}\,u^{\alpha}\,u^{\beta}.
\]
It follows that
\[
\nabla_n\Omega = \frac{1}{a}\,(\nabla_U\Omega - \Omega_{, \alpha}\,u^{\alpha}),
\]
can be calculated from the data given above.

\vspace{.2cm}

When starting from arbitrarily given standard Cauchy data $\hat{\delta}$ the functions $\Omega > 0$ and $\nabla_U\Omega < 0$  are not given but represent part of the conformal gauge freedom. Suitable choices will be discussed later.

\vspace{.2cm}

As a second step it will be convenient to derive all the unknowns entering the conformal field equations 
in a $g$-orthonormal frame $c_k$ on $S'$ which is adapted to $S'$ in the sense  that $c_0 = n$.
This frame, which is not needed in the final process, is introduced because it simplifies various discussions.
In a third step all the data will be expressed on $S'$ in terms of the 
$g$-orthonormal frame $e_k$ satisfying $e_0 = U$.

\vspace{.2cm}

To remove the gauge freedom in the transition $c_k \rightarrow e_k$,
we prescribe a specific field of Lorentz transformations $K^i\,_j$ on $S'$ which map the $g$-orthonormal  frame field $e_k$ with $e_0 = U$ onto a smooth  $g$-orthonormal frame $c_j = K^i\,_j\,e_i$ field 
with $c_0 = n$ by setting
 \begin{equation}
\label{K-def}
K^i\,_j 
=  \left(
\begin{array}{cc}
K^0\,_0,  & K^0\,_b\\
K^a\,_0, & K^a\,_b \\
\end{array}
\right)
= 
\left(
\begin{array}{cc}
- g(c_0, e_0)\,\,\,\,,  & g(c_0, e_b)  \\
\eta^{ad}\,g(c_0, e_d), & \delta^a\,_b + \frac{1}{1 - g(c_0, e_0)} \,\eta^{ad}\,g(c_0, e_d)\,g(c_0, e_b) \\
\end{array}
\right).
\end{equation}
In terms of the frame coefficients $e^{\mu}\,_k$ given by the solution $\Delta$ this reads
\[
K^i\,_j 
 = 
\left(
\begin{array}{cc}
\quad \quad a \quad \quad \,,  & - a\,e^0\,_b  \\
- a\,\eta^{ac}\,e^0\,_c\,, & \delta^a\,_b + \frac{a^2}{1 + a} \,\eta^{ac}\,e^0\,_c\,e^0\,_b \\
\end{array}
\right).
\]
It follows that  indeed 
\[
K^i\,_0\,e_i = K^0\,_0\,e_0 + K^a\,_0\,e_a =  - g(c_0, e_0)\,e_0 + \eta^{ad}\,g(c_0, e_d)\,e_a
= g(c_0, e_i)\,\eta^{ij}\,e_j = c_0.
\] 
In the following considerations  (\ref{2nd-a-form}) and
(\ref{2nd-n-form}) will be useful. A direct calculation verifies that $\eta_{ij}\,K^i\,_k\,K^j\,_l = \eta_{kl}$. 

The coefficients of the frame 
$c_k$ are given in the coordinates $x^{\mu}$ by
 \[
c ^{\mu}\,_k = \left(
\begin{array}{cc}
\,\,\,\,\,c^0\,_0\,\,\,\,\,\,,  & 0\\
c^{\alpha}\,_0\,, & c^{\alpha}\,_b \\
\end{array}
\right) = 
\left(
\begin{array}{cc}
\,\,\,\,\,\frac{1}{a}\,\,\,\,\,\,,  & 0\\
- \frac{1}{a}\,u^{\alpha}\,, & e^{\alpha}\,_b + \frac{1}{1 + a}\,u^{\alpha}\,e^0\,_b \\
\end{array}
\right), 
\]
and the coefficients of the 1-forms $\mu^k$ that satisfy 
$c^{\mu}\,_k \,\mu^k\,_{\nu} = \delta^{\mu}\,_{\nu}$ are so that 
 \[
\mu^{k}\,_{\nu} = \left(
\begin{array}{cc}
\mu^0\,_0\,,  & 0\\
\mu^{a}\,_0\,, & \mu^{a}\,_{\beta} \\
\end{array}
\right), 
\]
with
\[
\mu^0\,_0 = a, \quad \quad 
u^{\alpha} = (e^{\alpha}\,_b + \frac{1}{1 + a}\,u^{\alpha}\,e^0\,_b)\,\mu^b\,_0,
\]
\[
c^{\alpha}\,_b\,\mu^b\,_{\beta} = 
(e^{\alpha}\,_b + \frac{1}{1 + a}\,u^{\alpha}\,e^0\,_b)\,\mu^{b}\,_{\beta} = \delta^{\alpha}\,_{\beta},
\]
whence
\[
\mu^{a}\,_{\alpha}\,(e^{\alpha}\,_b + \frac{1}{1 + a}\,u^{\alpha}\,e^0\,_b)= \delta^{a}\,_{b},
\quad \quad 
\mu^{a}\,_{\alpha}\,u^{\alpha} = \mu^{a}\,_0.
\]
The comparison of 
\[
g = \eta_{jk}\,\mu^j\,\mu^k 
= - a^2\,d\tau^2 + \eta_{ab}\,\mu^a\,_{\alpha}\,\mu^b\,_{\beta}
(u^{\alpha}\,d\tau + dx^{\alpha})\,(u^{\beta}\,d\tau + dx^{\beta}),
\]
with the ADM representation 
\[
g = - (a\, d\tau)^2 + h_{\alpha \beta}\,(u^{\alpha}\,d\tau + dx^{\alpha})\,(u^{\beta}\,d\tau + dx^{\beta})
\]
gives then
\[
h_{\alpha \beta} =  \eta_{ab}\,\mu^a\,_{\alpha}\,\mu^b\,_{\beta}, 
\quad \quad 
a = \sqrt{1 + u^{\alpha}\,u^{\beta} \,h_{\alpha \beta}}.
\]
With the frame $c_j$ defined above we set  
\begin{equation}
\label{K-def}
M^i\,_j = 
\left(
\begin{array}{cc}
- g(e_0, c_0)\,\,\,\,,  & g(e_0, c_b)  \\
\eta^{ad}\,g(e_0, c_d), & \delta^a\,_b + \frac{1}{1 - g(e_0, c_0)} \,\eta^{ad}\,g(e_0, c_d)\,g(e_0, c_b) \\
\end{array}
\right).
\end{equation}
In terms of the frame coefficients $c^{\mu}_k$ this can be written
\[
M^i\,_j = 
\left(
\begin{array}{cc}
M^0\,_0,  & M^0\,_b\\
M^a\,_0, & M^a\,_b \\
\end{array}
\right)
= 
\left(
\begin{array}{cc}
\quad \quad a \quad \quad \,,  & u_{\alpha}\,c^{\alpha}\,_b  \\
\eta^{ac}\,u_{\alpha}\,c^{\alpha}\,_c\,, & \delta^a\,_b 
+ \frac{1}{1 + a} \,\eta^{ac}\,u_{\alpha}\,c^{\alpha}\,_c\,u_{\beta}\,c^{\beta}\,_b \\
\end{array}
\right)
\]

\[
= 
\left(
\begin{array}{cc}
\quad \quad a \quad \, ,& a\,e^0\,_b  \\
a\,\eta^{ac}\,e^0\,_c\,, & \delta^a\,_b + \frac{a^2}{1 + a} \,\eta^{ac}\,e^0\,_c\,e^0\,_b \\
\end{array}
\right).
\]
A direct calculation shows that $M^i\,_j\,K^j\,_k = \delta^i\,_k$
and $e_k = M^j\,_k\,c_j$.

\vspace{.2cm}

Because the fields $c_a$, $a = 1,2,3$ are tangential to $S'$
we can set 
\[
h'_{ab} = h_{\alpha \beta}\,c^{\alpha}\,_a\,c^{\beta}\,_b = \eta_{ab}, 
\quad \quad 
\kappa'_{ab} = \kappa_{\alpha \beta}\,c^{\alpha}\,_a\,c^{\beta}\,_b
\]
where here and in the following a prime is used to indicate when a tensor field is given in terms of the frame $c_k$. Directional derivatives with respect to $c_k$ will also indicated by a prime, so that
$\nabla'_k = \nabla_{c_k}$ etc.

\vspace{.2cm}

When the data for the conformal field equations are to be constructed by starting from standard Cauchy data, the frame $e_k$ is not available. Instead,  the frame $c_k$ has to be chosen first and $e_k$ will then be obtained by applying $M^i\,_k$. The
field $c_0$ is uniquely determined as the future directed unit normal to $S'$ but the frame $c_a$ tangent to $S'$ is only determined up to rotations. In the stability argument given below this freedom will have to be removed in a specific way.

\vspace{.2cm}

Connection coefficients with respect to the frame $c_k$ satisfying the relation 
$\nabla_{c_i} c_k = \gamma_i\,^j\,_k\,c_j$ with respect to the Levi-Civita connection $\nabla$ 
given by $g$
can only be defined if the frame is defined near 
$S'$. It will be convenient to extend the frame by the requirement  $\nabla_{c_0}c_k = 0$ and to define 
coordinates $\upsilon = x^{0'}$ and $x^{\alpha'}$
near $S'$ so that $x^{\mu'} = x^{\mu}$ on $S'$ and $<c_0, d\upsilon>\,=1$ and 
$<c_0,x^{\alpha'}>\, = 0$.
The coordinates $x^{\mu'}$ are then Gauss coordinates based on $S'$ and the coefficients 
$c^{\mu'}\,_k$ satisfy $c^{\mu'}\,_0 = \delta^{\mu'}\,_0$ and $c^{0'}\,_a = 0$ so that 
$c_0 = \partial_{\upsilon}$. The coordinates $x^{\mu}$ and $x^{\mu'}$ satisfy 
\[
\frac{\partial x^0}{\partial x^{0'}} = \,<n, d\tau> = \frac{1}{a}<U - u, d\tau>\, =  \frac{1}{a}, \quad
\frac{\partial x^0}{\partial x^{\alpha'}} = 0, 
\]
\[
\,\,\quad \quad \frac{\partial x^{\alpha}}{\partial x^{0'}} =  \frac{1}{a}<U - u, d x^{\alpha}>\, = -  \frac{1}{a}\,u^{\alpha}
,\quad 
\frac{\partial x^{\alpha}}{\partial x^{\alpha'}} = \delta^{\alpha}\,_{\alpha'}
\quad \mbox{on $S'$}, 
\]
so that the relation $e^{\mu'}\,_k = M^j\,_k\,c^{\mu'}\,_j$ can be used to determine on $S'$
\[
e^{\mu}\,_k = \frac{\partial x^{\mu}}{\partial x^{\mu'}}\,c^{\mu'}\,_l\,M^l\,_k.
\] 
The connection coefficients with respect to $c_k$ can now be defined. They satisfy 
\[
\gamma_0\,^j\,_k = 0, \quad \gamma_a\,^0\,_b = \kappa'_{ab} = \kappa'_{ba},\quad
\gamma_a\,^c\,_0 = \kappa'_{ab}\,h'^{bc}, \quad
\gamma_a\,^d\,_b\,c_d = D_{c_a} c_b \quad \mbox{on $S'$},
\]
where $D$ denotes the Levi-Civita connection of the metric $h$ on $S'$. 

The connection coefficients in the frame $c_k$ are related to the connection coefficients in the frame $e_k$ by
\[
\Gamma_i\,^j\,_k 
= K^j\,_n
\left(M^n\,_{k, \,\mu'}\,e^{\mu'}\,_i + \gamma_l\,^n\,_p\,M^l\,_i\,M^p\,_k \right) 
\]
\[
= K^j\,_n
\left(M^n\,_{k, \,0'}\,e^{0'}\,_i + M^n\,_{k, \,\alpha'}\,e^{\alpha'}\,_i 
 + \gamma_l\,^n\,_p\,M^l\,_i\,M^p\,_k \right).
\]
Apart from $M^n\,_{k, \,0'}$, which can only be determined by taking into account  the evolution equations for the frame $e_k$, 
all the other terms in the expression above can be calculated from the data available so far. 
The relation $e^{\mu'}\,_k = M^j\,_k\,c^{\mu'}\,_j$ implies
\[
e^{\mu'}\,_{k,\,0'} = M^j\,_{k,\,0'}\,c^{\mu'}\,_j  + M^j\,_k\,c^{\mu'}\,_{j,\,0'}.
\]
The first structural equation with respect to the frame $c_k$ gives
\[
c^{\mu'}\,_{j,\,0'} = \delta^{\mu'}\,_{\alpha'}\,\delta^a\,_j\,c^{\alpha'}\,_{a,\,0'}
= - \delta^{\mu'}\,_{\alpha'}\,\delta^a\,_j\,\gamma_a\,^b\,_0\,c^{\alpha'}\,_b
= - \delta^{\mu'}\,_{\alpha'}\,\delta^a\,_j\,\kappa'_{ac}\,h'^{bc}\,c^{\alpha'}\,_b
\quad \mbox{on $S'$}.
\]
The field $e_0 = U = U'^k\,c_k$, given on $S'$ by $U = a\,c_0 + u'^a\,c_a$
with $u'^a = \mu^a\,_{\alpha'}\,u^{\alpha'}$, must thus satisfy by  (\ref{first-g-conf-geod-equ})
\[
0 = U'^k\,_{, \,\mu'}\,c ^{\mu'}\,_l\,U'^l + U'^l\,U'^j\,\gamma_l\, ^k\,_j
+  <U,f>U'^k + f'^k \quad \quad
\]
\[
\quad  = a\,U'^k\,_{, \,0'} + U'^k\,_{, \,\alpha'}\,u ^{\alpha'}
+ U'^l\,U'^j\,\gamma_l\, ^k\,_j +  <U,f>U'^k + f'^k \quad \mbox{on $S'$}.
\]
The fields $e_a = e'^k\,_a\,c_k$ must satisfy $\mathbb{F}_U e_a = 0$, which implies with 
(\ref{first-g-conf-geod-equ}) 
\[
0 = a\,e'^k\,_{c,\,0'} + e'^k\,_{c,\,\alpha'}u^{\alpha'} +U'^i\,e'^j\,_c\gamma_i\,^k\,_j 
+ f'_l\,e'^l\,_c\,U'^k - U'_l\,e'^l\,_c\,f'^k \quad \mbox{on $S'$}.
\]
These relations determine $c^{\mu'}\,_{j,\,0'}$, $e^{\mu'}\,_{k,\,0'}$ whence
$M^j\,_{k,\,0'}$ and $\Gamma_i\,^j\,_k $ uniquely from the given data on $S'$ once $f'_k$ is 
given there.

\vspace{.2cm}

Our gauge requires that the tensorial field 
\[
N'_k =    \nabla'_k\Omega  + (\nabla_U\Omega +
\Omega <U, f>)\,U'_k +  \Omega\,f'_k,
\]
vanishes on $S'$. The condition that its orthogonal projection $N'_a$ into $S'$ vanishes gives
\[
f'_a = - \frac{1}{\Omega}\,\{ \nabla'_a\Omega  + (\nabla_U\Omega +
\Omega <U, f>)\,u'_a\} \quad \mbox{on $S'$}.
\]
If this is satisfied it follows with $U_k = U'_i\,\,M^i\,_k$, $N_k = N'_i\,\,M^i\,_k$
\[
0 = U^k\,N_k = U'^k\,N'_k = a\,n'^k\,N'_k,
\]
and thus together $N'_k = 0$. The relation 
\[
f'_0 = n'^k\,f'_k = \frac{1}{a}\,(<U, f> - u'^a\,f'_a),
\]
shows that $f'_k$ is determined from the data given on $S'$ only up to $f_0 = \,<U, f>$. This is consistent with the fact remarked on earlier
that the quantity $f_0$  is pure gauge and can be chosen arbitrarily. With a  suitable choice of $f_0$
(made in a specific way  later) we can the set $f_k = f'_j\,\,M^j\,_k$.

\vspace{.2cm}

The Einstein equations and the conformal rescaling of the density  imply 
$R[\hat{g}] = 4\,\lambda + \Omega^3\rho$. With this the conformal transformation law of the Ricci scalar gives
\[
\nabla_{\mu}\nabla^{\mu}\Omega + \frac{1}{6}\,R[g]\,\Omega 
= \frac{2}{\Omega}\,\nabla_{\mu}\Omega\,\nabla^{\mu}\Omega 
+ \frac{1}{6\,\Omega}\,R[\hat{g}]
=  \frac{2}{\Omega}\,\nabla_{\mu}\Omega\,\nabla^{\mu}\Omega 
+ \frac{1}{6\,\Omega}\,(4\,\lambda + \Omega^3\rho).
\]
With the gauge condition $R]g| = 0$ we thus set
\[
4\,s = \nabla'_k\nabla'^k\Omega =  \frac{2}{\Omega}\,\nabla'_{i}\Omega\,\nabla'^{i}\Omega 
+ \frac{1}{6\,\Omega}\,(4\,\lambda + \Omega^3\rho).
\]
The second equation determines $\partial^2_{\upsilon}\,\Omega = c_0(c_0\,\Omega)$
in terms of known data because
\[
 \nabla'_k\nabla'^k\Omega = - \nabla'_0\nabla'_0\Omega  +  \eta^{ab}\,\nabla'_a\nabla'_b\Omega 
  = - c_0(c_0\,\Omega)
+ \eta^{ab} (D'_aD'_b\Omega - \kappa'_{ab}\,\nabla_n\Omega)
\quad \mbox{ on $S'$}.
\]
Thus $s$ and $\nabla'_j\nabla'_k\Omega$ are determined on $S'$ from known data and the scalar equation (\ref{f-alg-equ})
is satisfied there. Given $s$ and $\chi_{ab} = \Gamma_a\,^0\,_b$, the fields
$\zeta_{ab}$ and $\xi$ are then defined on $S'$ by (\ref{regularizing-unknowns}).

\vspace{.2cm}

The conformal transformation law of the Schouten tensor, the field equations, and the conformal rescalings of the flow vector field and the density give
\[
L_{\mu\nu} = \hat{L}_{\mu\nu} - \frac{1}{\Omega}\,\nabla_{\mu}\,\nabla_{\nu}\Omega
+ \frac{1}{2\,\Omega^2}\,\nabla_{\rho}\Omega\,\nabla^{\rho}\Omega\,g_{\mu\nu}
\]
\[
= \frac{1}{6}\,\lambda\,\Omega^{-2}\,g_{\mu\nu} 
+ \Omega\, \rho\left(\frac{1}{2}\,U_{\mu}\,U_{\nu} + \frac{1}{6}\,g_{\mu\nu}\right)
- \frac{1}{\Omega}\,\nabla_{\mu}\,\nabla_{\nu}\Omega
+ \frac{1}{2\,\Omega^2}\,\nabla_{\rho}\Omega\,\nabla^{\rho}\Omega\,g_{\mu\nu},
\]
and we set 
\[
L'_{i j} = 
\frac{1}{6}\,\lambda\,\Omega^{-2}\,g'_{i j} 
+ \Omega\,\rho\left(\frac{1}{2}\,U'_{i}\,U'_{j} + \frac{1}{6}\,g'_{i j}\right)
- \frac{1}{\Omega}\,\nabla'_{i}\,\nabla'_{j}\Omega
+ \frac{1}{2\,\Omega^2}\,\nabla'_{l}\Omega\,\nabla'^{l}\Omega\,g'_{i j} \quad \mbox{ on $S'$ }.
\]
By the way $\nabla'_0\nabla'_0\Omega$ has been determined above it follows that 
$g'^{ik}\,L'_{ik} = \frac{1}{6}\,R[g] = 0$. The appropriate data on $S'$ for the reduced field equations 
are then given by $L_{jk} = L'_{il}\,\,M^i\,_j\,M^l\,_k$.

\vspace{.2cm}

To determine the rescaled conformal Wey tensor we observe that the Gauss and the Codazzi equation 
with respect to $S'$ read in terms of the frame $c_k$}
\[
R'_{abcd}[g] = R'_{abcd}[h] + \kappa'_{a c}\,\kappa'_{b d} - \kappa'_{a d}\,\kappa'_{b c},
\]
\[
n'^kR'_{kabc}[g] = D'_{c} \kappa'_{b a} - D'_{c}\,\kappa'_{d a},
\]
where the fields on the right hand sides can be determined from the data available so far. With 
$L'_{jk}$ as given above, the general relation 
\[
R'_{i j k l}[g] =
2\,\{g'_{i[k}\,L'_{l] j}  +
\,L'_{i[k}\,g'_{l] j}\} 
+ C'_{i j k l}, 
\]
then allows us to calculate the components 
$C'_{abcd}[g]$ and $n'^kC'_{kabc}[g]$ of the conformal Weyl tensor.
The conformal Weyl
tensor admits the decomposition 
\[
C'_{ijkl} = 2 \left( k'_{i[k}\,e'_{l]j}  - k'_{j[k}\,e'_{l]i} 
+ n'_{[k}\,m'_{l]m}\,\epsilon'^{m}\,_{ij} 
+ n'_{[i}\,m'_{j]m}\,\epsilon'^{m}\,_{kl} \right).
\] 
where $h'_{jk} = g'_{jk} + n'_j\,n'_k$ and $k'_{jk} = g'_{jk} + 2\,n'_j\,n'_k$ and 
$e'_{ik} = h'_i\,^m\,h'_k\,^n\,C'_{mjnl}\,n'^j\,n'^l$ and 
$m'_{ik} = h'_i\,^m\,h'_k\,^n\,C'^*_{mjnl}\,n^j\,n^l$ with
$C'^*_{ijkl} = \frac{1}{2}\,C'_{ijmn}\,\epsilon'^{mn}\,_{kl}$
denote the electric and magnetic part of the conformal Weyl tensor 
{\it with respect to} $n$ in the frame $c_k$ respectively. 
It holds $e'_{ij} = e'_{ji}$, $e'_{ij}\,n'^j = 0$, $e'_i\,^i = 0$ and similar 
relations hold for $m'_{ij}$. 

It follows that 
\[
C'_{abcd} = 2\,( h'_{a[c}\,e'_{d]b} + e'_{a[c}\,h'_{d]b})
\quad \mbox{whence} \quad  
e'_{bd} = h'^{ac}\,C'_{abcd},
\] 
and 
\[
n'^k C'_{kbcd} = 2\,(n'_{[i}\,m'_{j]m}\,\epsilon'^{m}\,_{kl})
\quad \mbox{whence} \quad  
m'_{ab} = - \frac{1}{2}\,n'^k\,C'_{kbcd}\,\epsilon'_{b}\,^{cd}. 
\]
The tensors $C'_{ijkl}$ and $W'_{ijkl}  = \Omega^{-1}\,C'_{ijkl}$ whence
$W_{ijkl} = W'_{mnpq} \,M^m\,_i\,M^n\,_j\,M^p\,_k\,M^q\,_l$ can thus be determined from the given data and thus also $U$-electric and -magnetic parts $w_{ij}$ and $w^*_{kl}$ of 
$W_{ijkl}$ which enter the reduced conformal field equations.

\vspace{.2cm}

The conformal field equations and their unknowns are derived from the Einstein equations by conformal rescalings, the use of various differential identities, and the use of the frame formalism.
This leaves a coordinate, frame, and conformal gauge freedom which is controlled by suitable initial data and propagation laws for the coordinates, the frame field, and the conformal factor
(controlled here implicitly by the requirement $R[g]= 0$). Following this procedure it follows from the discussion above how to derive from a given smooth solution 
$\hat{\delta} = (\hat{h}_{\alpha \beta}, \,\hat{\kappa}_{\alpha \beta}, \,\hat{u}^{\alpha}, \,\hat{\rho})$
to the constraints (\ref{hat-Ham-constr}) and (\ref{hat-mom-constr}) 
and given smooth gauge dependent  fields
\begin{equation}
\label{gauge-rep-fields-on-S'}
\Omega > 0,  \,\,\,  \nabla_U\Omega < 0,  \,\,\,  f_0 = \,<U, f>,  \,\,\, \mbox{and a smooth
$h$-orthonormal  field $c_a$ on $S'$},
\end{equation}
the unknowns $\Delta'_{S'}$ on $S'$ of the conformal field equations in the frame $c_k$ and also the unknowns 
\begin{equation}
\label{conf-e-data-onS'}
\Delta_{S'} 
= (e^{\mu}\,_k, \,\,\, \Gamma_i\,^j\,_k,  \,\,\,  \zeta_{ab}, \,\,\, \xi,  \,\,\,
f_k,  \,\,\,  \Omega,  \,\,\,  \nabla_j\Omega,  
 \,\,\,  s,  \,\,\, L_{jk},  \,\,\,  W^i\,_{jkl},  \,\,\, 
U^k,  \,\,\,  \rho), 
\end{equation}
in the frame $e_k$ on $S'$. 

Written in terms of the frame $c_k$ and the frame coefficients $c^{\mu'}\,_k$ as defined above, the conformal field equations allow us to derive from the data $\Delta'_{S'}$ a formal expansion type solution in terms of the coordinate $\upsilon$ so that the complete set of conformal field equations is satisfied at all orders. The constraints are satisfied because of differential identities and the fact that the data $\hat{\delta}$ satisfy the `physical' constraints. 

A similar formal expansion is obtained in terms of the coordinate $\tau$ if the equations and the data are expressed in terms of the frame $e_k$.
In this case the expansion coefficients are seen, however, to be the coefficients of a Taylor expansion of an actual smooth solution to the conformal field equations because the equations comprise the hyperbolic system of reduced conformal field equations. 

The life time of the solution in the given gauge depends, of course, on the data (\ref{conf-e-data-onS'}) and in particular on the 
choice of the free fields in (\ref{gauge-rep-fields-on-S'}). Suppose 
\begin{equation}
\label{comparison-solution}
\Delta^{\star}(\tau) 
= (e^{\star\,\mu}\,_k, \,\,\, \Gamma^{\star}_i\,^j\,_k,  \,\,\,  \zeta^{\star}_{ab}, \,\,\, \xi^{\star},  \,\,\,
f^{\star}_k,  \,\,\,  \Omega^{\star},  \,\,\,  \nabla_j\Omega^{\star},  
 \,\,\,  s^{\star},  \,\,\, L^{\star}_{jk},  \,\,\,  W^{\star i}\,_{jkl},  \,\,\, 
U^{\star k},  \,\,\,  \rho^{\star}), 
\end{equation}
is one of the solutions to the conformal field equations considered in the previous subsection.
It exists and is smooth for 
$\tau_* \le \tau \le \tau_{**}$ with $ \Omega^{\star} \rightarrow 0$ as $\tau \rightarrow 0$
so that $S_0$ represents the conformal boundary at future time-like infinity for the physical solution associated with 
$\Delta^{\star}(\tau)$. Denote by
$\Delta^{\star}_{S'} = \Delta^{\star}(\tau_*)$ the data for the reduced equations on $S'$ and by  
$\hat{\delta}^{\star} = (\hat{h}^{\star}_{\alpha \beta}, \,\hat{\kappa}^{\star}_{\alpha \beta}, 
\,\hat{u}^{\star \alpha}, \,\hat{\rho}^{\star})$ the physical data induced by this solution on $S'$. 
Let $\hat{\delta} = (S',\,\hat{h}_{\alpha \beta}, \,\hat{\kappa}_{\alpha \beta}, \,\hat{u}^{\alpha}, \,\hat{\rho})$ denote a smooth solution 
to the constraints (\ref{hat-Ham-constr}) and (\ref{hat-mom-constr}),  $\Delta_{S'}$ the corresponding initial data on $S'$ for the reduced conformal field equations as considered  in (\ref{gauge-rep-fields-on-S'}), and $\Delta(\tau)$, where 
$\tau \in  [\tau_*, \tau_* + \tau^*[$ with some   $ \tau^*  > 0$, the solution to the conformal field equations determined by these data.

To compare the life times of the solutions $\Delta^{\star}(\tau)$ and $\Delta(\tau)$ the corresponding gauge conditions must be comparable. It will be assumed that the data $\Delta_{S'}$ have been constructed such that
\[
\Omega = \Omega^{\star}, \quad \nabla_U\Omega = \nabla_U\Omega^{\star}, \quad 
f_0 = f^{\star}_0 \quad \mbox{on $S'$}.
\]
Let $h^{\star}_{\alpha \beta}  =  \Omega^{\star 2} \,\hat{h}^{\star}_{\alpha \beta}$,
and $h_{\alpha \beta} =  \Omega^{2} \,\hat{h}_{\alpha \beta} = \Omega^{\star 2} \,\hat{h}_{\alpha \beta}$
denote the metric induced on $S'$ by the solution $\Delta^{\star}(\tau)$ and  $\Delta(\tau)$ respectively. 
As discussed above, the frame $e^{\star}_k$ given by  the data  $\Delta^{\star}_{S'}$ can be used to define a field of Lorentz transformation $K^{\star j}\,_l$ on  $S'$ so that the relation $c^{\star}_k = K^{\star j}\,_k\,e^{\star}_j$ defines a frame field on $S'$ 
for which $c^{\star}_0$ is normal to $S'$. The  fields $c^{\star}_a$, $a = 1, 2, 3$, then define an $h^{\star}$-orthonormal frame
field on $S'$. It will be assumed in the following that the $h$-orthonormal  field $c_a$ has been chosen so
that $c_a = c^{\star}_c\,\alpha^c\,_a$ with a $3 \times 3$ matrix $\alpha^c\,_a$ that satisfies
$\alpha^1\,_1 > 0$, $\alpha^2\,_2 > 0$, $\alpha^3\,_3 > 0$, and  $\alpha^c\,_a = 0$ if $a < c$.
The frame $c_a$ so defined is smooth and fixed uniquely so that 
$\alpha^c\,_a \rightarrow \delta^a\,_c$ precisely if  $c_a \rightarrow c^{\star}_a$.

The point of these choices is that the
space-time conditions $R[g^{\star}] = 0$ and $R[g] = 0$ combine with these gauge conditions on $S'$ to ensure that
$||\hat{\delta} - \hat{\delta}^{\star} || \rightarrow 0$ if and only if $|||\Delta_{S'} - \Delta^{\star}_{S'}||| \rightarrow 0$, 
where the norms are meant to indicate Sobolev norms on $S'$ which are chosen corresponding to the differentiability order of the fields involved.

\vspace{.2cm}

We can invoke now the Cauchy stability property which holds for  hyperbolic equations to conclude that  
for data $\hat{\delta}$ sufficiently close to $\hat{\delta}^{\star}$ or, equivalently, 
for data $\Delta_{S'}$ sufficiently close to $\Delta^{\star}_{S'}$ the solution $\Delta(\tau)$ of the conformal field equations
that develops from the data $\Delta_{S'}$ also exists in the interval  $\tau_* \le \tau \le \tau_{**}$
and the conformal factor $\Omega$ supplied by $\Delta(\tau)$ is negative on $S_{\tau_{**}}$ \cite{kato}. 
This conclusion may require repeated patchings (see \cite{friedrich:1991}).

There exists then a map $S \ni q \rightarrow \tau(q) \in ]\tau_*, \tau_{**}[$ so that
$\Omega(\tau(q) , q) = 0$ for $q \in S$ 
and  $\Omega(\tau, q) > 0$ if $\tau_* \le \tau < \tau(q)$.
Equation (\ref{f-alg-equ}) then implies that on the subset 
${\cal J}^+ = \{(\tau(q), q), q \in S\}$ of $\mathbb{R} \times S$ the gradient
$\nabla^{i}\Omega$ is time-like 
for the metric $g$ supplied by $\Delta(\tau)$. It follows that ${\cal J}^+$ defines a smooth space-like hypersurface 
which represents a conformal boundary in the infinite future of  the set 
$\hat{M} = \{(\tau, q) \in \mathbb{R} \times S \,|\,\tau_* \le \tau < \tau(q)\}$ on which the fields
$\hat{g}_{\mu\nu} = \Omega^{-2}\,g_{\mu \nu}$, $\hat{U}_{\mu} = \Omega^{-1}\,U^{\mu}$, $\hat{\rho}' = \Omega^3\,\rho$ define a 
 smooth solution to the Einstein-$\lambda$-dust equations. The smooth asymptotic end data induced by its conformal extension
$\Delta(\tau)$ on ${\cal J}^+ \sim S$ belongs then to the class of conformal end data considered in section 
\ref{as-end-dat}.
Combining the results of the last two subsection we obtain Theorem \ref{main-result}.

\vspace{.5cm}

\noindent
{\bf Acknowledgements}: I would like to thank the relativity group at Cordoba in Argentina, where this work was begun,
for hospitality and discussions.

}


\begin{thebibliography}{9}

\bibitem{bartnik:isenberg}
R. Bartnik, J. Isenberg.
\newblock The constraint equations.
\newblock In: P. T. Chru\'sciel, H. Friedrich (eds.):
{\it The Einstein equations and the large scale behaviour of gravitational fields.}
\newblock Birkh\"auser, Basel, 2004.

\bibitem{beyer:2008}
F. Beyer.
\newblock Investigations of solutions of Einstein's field equations close to lambda-Taub-NUT.
\newblock {\it Class. Quantum Gravity} 25 (2008) 235005.

\bibitem{friedrich:hyperbolic}
H. Friedrich.
\newblock On the hyperbolicity of Einstein's and other gauge field equations.
\newblock {\em Commun. Math. Phys.} 100 (1985) 525 - 543.

\bibitem{friedrich:1986a}
H. Friedrich.
\newblock Existence and structure of past asymptotically simple solution of 
Einstein's field equations with positive cosmological constant.
\newblock {\it J. Geom. Phys.} 3 (1986) 101 - 117.

\bibitem{friedrich:1986b}
H. Friedrich.
\newblock On the existence of n-geodesically complete or future 
complete solutions of Einstein's field equations with smooth asymptotic structure.
\newblock {\it Commun. Math. Phys.} 107 (1986), 587 - 609.

\bibitem{friedrich:1991}
H. Friedrich.
\newblock On the global existence and the asymptotic behaviour of solutions to the Einstein-Maxwell-Yang-Mills equations.
\newblock {\it J. Differential Geometry} 34 (1991) 275 - 345.

\bibitem{friedrich:AdS}
H. Friedrich.
\newblock Einstein equations and conformal structure: existence of
anti-de Sitter-type space-times.
\newblock { \it J. Geom. Phys.} 17 (1995) 125--184.

\bibitem{friedrich98}
Friedrich, H. (1998) 
\newblock Evolution equations for gravitating ideal fluid
bodies in general relativity. 
\newblock {\it Phys. Rev.} D 57, 2317--2322.

\bibitem{friedrich:cg on vac}
H. Friedrich.
\newblock Conformal geodesics on vacuum space-times.
\newblock {\em Commun. Math. Phys.} 235 (2003) 513 - 543. 

\bibitem{friedrich:massive fields}
H. Friedrich.
\newblock Smooth non-zero rest-mass  evolution across time-like infinity.
\newblock {\em Ann. Henri Poincar\'e} 16 (2015) 2215 - 2238.

\bibitem{friedrich:beyond:2015}
H. Friedrich.
\newblock  Geometric asymptotics and beyond.
\newblock  In: L.Bieri, S.-T. Yau (eds), {\em Surveys in Differential Geometry}, Vol.20.
\newblock International Press, Boston, 2015.
\newblock arXiv:1411.3854 

\bibitem{friedrich:rendall}
H. Friedrich, A. Rendall.
\newblock The Cauchy Problem for the Einstein Equations.
\newblock In: B. Schmidt (ed.): {\it Einstein's Field Equations and
Their Physical Implications.}
\newblock Springer, Lecture Notes in Physics, Berlin 2000.

\bibitem{hadzic-speck-2015}
M. Had$\check{z}$i\'c, J. Speck.
\newblock The global future stability of the FLRW solutions to the Dust--Einstein system with a positive cosmological constant.
\newblock {J. Hyperbolic Differential Equations} 12 (2015) 87.

\bibitem{hawking:ellis}
S. Hawking, G. Ellis.
\newblock {\em The large scale structure of space-time}.
\newblock Cambridge University Press, 1973.

\bibitem{kato}
T. Kato.
\newblock The Cauchy problem for quasi-linear symmetric hyperbolic systems.
\newblock {\em Arch. Ration. Mech. Anal.}, 58 (1975) 181 - 205.

\bibitem{luebbe:valiente-kroon:2013}
C. L\"ubbe, J. A. Valiente Kroon.
\newblock A conformal approach to the analysis of the non-linear stability of radiation cosmologies.
\newblock {\it Ann. Phys.} 328 (2013) 1.
 
\bibitem{penrose:1965}
R. Penrose.
\newblock Zero rest-mass fields including gravitation: asymptotic behaviour.
\newblock {\it Proc. Roy. Soc. Lond.} A 284 (1965) 159 - 203.

\bibitem{penrose:2011}
R. Penrose.
\newblock {\it Cycles of Time}.
\newblock  Vintage, London, 2011.

\bibitem{ringstroem:2008}
H. Ringstr\"om.
\newblock Future stability of the Einstein-non-linear scalar field system.
\noindent {\it Invent. math.} 173 (2008) 123 - 208.




\end{thebibliography}
\end{document}